\documentclass{article}
\usepackage{amsmath}
\usepackage{graphicx}
\usepackage{caption}
\usepackage{algorithm}
\usepackage{algorithmic}
\usepackage{caption}
\usepackage{subcaption}
\usepackage{booktabs} 
\usepackage{tikz-cd}
\usetikzlibrary{fit}
\usetikzlibrary{arrows}
\usepackage{comment}
\usepackage{amsthm}
\usepackage{amssymb}
\usepackage{url}
\usepackage{tikz}
\usepackage{tikz-qtree,tikz-qtree-compat}
\usetikzlibrary{positioning}
\usepackage{wrapfig}
\usetikzlibrary{positioning,calc}
\theoremstyle{definition}
\newtheorem{definition}{Definition}
\newtheorem{theorem}{Theorem}
\newtheorem{lemma}{Lemma}
\usepackage{enumitem}
\usepackage[preprint]{neurips_2025}
\usepackage{subcaption}

\usepackage[utf8]{inputenc} 
\usepackage[T1]{fontenc}    
\usepackage{hyperref}       
\usepackage{url}            
\usepackage{booktabs}       
\usepackage{amsfonts}       
\usepackage{nicefrac}       
\usepackage{microtype}      
\usepackage{xcolor}         

\title{Recover Experimental Data with Selection Bias using Counterfactual Logic}

\author{
  Jingyang He, \;\; Shuai Wang, \;\; Ang Li \\
  Department of Computer Science \\ 
  Florida State University \\ 
  Tallahassee, FL 32306 \\
  \texttt{\{jh24o,\,sw23vf,\,al23bp\}@fsu.edu}
}

\begin{document}

\maketitle

\begin{abstract}

Selection bias, arising from the systematic inclusion or exclusion of certain samples, poses a significant challenge to the validity of causal inference. While \citet{bareinboim2022recovering} introduced methods for recovering unbiased observational and interventional distributions from biased data using partial external information, the complexity of the backdoor adjustment and the method’s strong reliance on observational data limit its applicability in many practical settings. In this paper, we formally discover the recoverability of \(P(Y^*_{X^*})\) under selection bias with experimental data. By explicitly constructing counterfactual worlds via Structural Causal Models (SCMs), we analyze how selection mechanisms in the observational world propagate to the counterfactual domain. We derive a complete set of graphical and theoretical criteria to determine that the experimental distribution remain unaffected by selection bias. Furthermore, we propose principled methods for leveraging partially unbiased observational data to recover \(P(Y^*_{X^*})\) from biased experimental datasets. Simulation studies replicating realistic research scenarios demonstrate the practical utility of our approach, offering concrete guidance for mitigating selection bias in applied causal inference.
\end{abstract}

\section{Introduction}
\label{inrto}


Selection bias (\citet{heckman1979sample}) arises when the analyzed sample systematically fails to represent the target population due to a non-random selection mechanism. Typically driven by unobserved factors that influence both sample inclusion and outcomes, selection bias distorts observed associations and obscures true treatment effects, thereby critically undermining the validity of causal estimations across all causal layers defined by \cite{pearl2009causality}. Even randomized controlled trials within a selected subgroup cannot fully eliminate such bias, as entry into the subgroup is itself governed by a selection mechanism. For instance, researchers may preferentially recruit patients with severe or complex conditions to test a novel targeted therapy, neglecting those with mild symptoms. As a result, any inference regarding probabilities of causation (i.e., counterfactuals) (\citet{balke1994counterfactual}) based on this subgroup systematically deviates from reality.

Such preferential selection poses challenges to inference in many domains, including epidemiology(\citet{enzenbach2019evaluating}, \citet{millard2023exploring}), artificial intelligence(\citet{schnabel2016recommendations}, \citet{huang2022different}, \citet{gururangan2018annotation}, \citet{geva2019we}), economics(\citet{lalonde1986evaluating}, ), and even the hottest large language models(\citet{bender2021dangers},\citet{manela2021stereotype}, \citet{mcmilin2022selection}). More fundamentally, selection bias undermines the foundations of causal and statistical inference, rendering advanced causal estimands and statistical measures, such as the Effect of Treatment on the Treated (\citet{rubin1974estimating}), Probability of Necessity, Probability of Sufficiency, and Probability of Necessity and Sufficiency (\citet{pearl2022probabilities}) derived from biased datasets inherently unreliable.

Over the past years, substantial advances have been made in correcting selection bias from a causal standpoint. \citet{bareinboim2022recovering} introduced the rigorous theory of s‑recoverability, precisely characterizing the graphical and algebraic conditions under which biased observational and interventional distributions can be re‑weighted to recover unbiased causal estimands via integration over selected subpopulations. \citet{rosenbaum1983central}'s propensity score theory provides a theoretically rigorous and practically implementable framework for recovering unbiased average causal effects, such as the Average Treatment Effect (ATE) and the Average Treatment Effect on the Treated (ATT) (\citet{rubin1974estimating}), from non-randomized observational data. Moreover, the “Graphical Models for Inference with Missing Data” framework introduced by \citet{mohan2013graphical} can likewise be viewed as an alternative approach to modeling selection bias.

The selection-backdoor adjustment, as proposed by \citet{bareinboim2022recovering}, allows for identifying the  distribution \(P(Y_x)\) using a combination of an unbiased observational distribution \(P(Z)\) and a biased observational distribution $P(Y|x, \textbf{z}, S=1)$. However, in practice, observational data \(P(Y | x, \textbf{z}, S=1)\) are often not more accessible than experimental data. For instance, after a new drug is released, it may be difficult to obtain sufficient observational follow-up, leaving only biased experimental data available. In such cases, the assumptions required by selection-backdoor adjustment break down. To overcome this limitation, we turn to counterfactual reasoning via a twin-network formulation, which enables the recovery of the unbiased distribution \(P(Y_x)\) based on biased experimental data, without requiring access to observational distributions \(P(Y | x, \textbf{z}, S=1)\).

In summary, our paper makes the following key contributions:
\paragraph*{Contributions:}
\begin{itemize}[leftmargin=*]
  \item \textbf{Nonparametric recoverability criterion and theorem.} We introduce a nonparametric definition and theorem that exactly characterize whether selection bias perturbs the distribution \(P(Y^*_{X^*})\), and how to recover distribution \(P(Y^*_{X^*})\) using biased experimental data.
  \item \textbf{Twin-Network recoverability framework.} Leveraging Pearl’s twin‑network, we decouple identification from recovery process and reconstruct an unbiased \(P(Y_x)\) from biased subgroup data and external distribution \(P(Z)\).
  \item \textbf{Scalable validation.} Extensive simulations (varying sample size and seeds) show rapid convergence to ground truth and large reductions in bias across multiple metrics.
\end{itemize}

\section{Preliminary}
\label{gen_inst}

\subsection{Modeling Selection Bias in Causal Graphs}

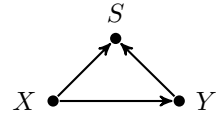
\begin{wrapfigure}{r}{0.3\textwidth}
  \vspace{-1em} %
  \centering
  \begin{tikzpicture}[->,>=stealth',node distance=2cm,
  thick,main node/.style={circle,fill,inner sep=1.5pt}]
  \node[main node] (0) [label=above:{\(S\)}]{};
  \node[main node] (1) [below left = 1cm of 0,label=left:\(X\)]{};
  \node[main node] (2) [right =1.5cm of 1,label=right:\(Y\)] {};
  \path[every node/.style={font=\sffamily\small}]
    (1) edge node {} (2)
    (1) edge node {} (0)
    (2) edge node {} (0);
\end{tikzpicture}
\caption{Causal graph with selection node}
\label{fig:causalg1}
\end{wrapfigure}

As shown in Figure~\ref{fig:causalg1}, \citet{bareinboim2012controlling} introduced an explicit selection node S into the underlying causal DAG, with directed edges from all variables hypothesized to influence sample inclusion (e.g. Treatment \(X\) and Outcome \(Y\)). The node \(S\) is a binary indicator representing sample inclusion, with \(S=1\) denoting a selected sample and \(S=0\) denoting exclusion. Directed edges clearly illustrate which nodes influence sample selection. In this study, we adopt this foundational setup and extend it explicitly into the counterfactual domain.

\subsection{Twin network}
\begin{figure}[h]
\centering
\resizebox{1.0\textwidth}{!}{%
    \begin{subfigure}[b]{0.35\textwidth}
    \centering
    \begin{tikzpicture}[->,>=stealth',node distance=2cm,
      thick,main node/.style={circle,fill,inner sep=1.5pt}]
      \node[main node] (0) [label=above:{\(S\)}]{};
      \node[main node] (1) [above left = 1cm of 0,label=left:\(X\)]{};
      \node[main node] (2) [below =1.5cm of 1,label=below:\(Y\)] {};
      \node[main node] (3) [above right = 0.6cm of 0, label=above:{\(U_S\)}]{};
      \node[main node] (4) [below = 0.8 cm of 3, label=above:{\(U_Y\)}]{};
      \node[main node] (5) [above right = 0.5 cm of 1, label=above:{\(U_X\)}]{};
      \node[main node] (6) [below right = 0.6cm of 3, label=above:{\(S^*\)}]{};
      \node[main node] (7) [above right = 1cm of 6,label=right:\(X^*\)]{};
      \node[main node] (8) [below = 1.5cm of 7, label=below:{\(Y^*\)}]{};
      \path[every node/.style={font=\sffamily\small}]
        (1) edge node {} (2)
        (1) edge node {} (0)
        (7) edge node {} (6)
        (7) edge node {} (8);
      \draw [dashed] (3) -- (0);
      \draw [dashed] (4) -- (2);
      \draw [dashed] (3) -- (6);
      \draw [dashed] (5) -- (1);
      \draw [dashed] (4) -- (8);
    \end{tikzpicture}
    \caption{}
    \label{causalg_a}
    \end{subfigure}\hfill
    \begin{subfigure}[b]{0.35\textwidth}
    \centering
    \begin{tikzpicture}[->,>=stealth',node distance=2cm,
      thick,main node/.style={circle,fill,inner sep=1.5pt}]
      \node[main node] (0) [label=left:{\(X\)}]{};
      \node[main node] (1) [below =1.5cm of 0,label=below:\(Y\)] {};
      \node[main node] (2) [right = 0.5cm of 0, label=above:{\(S\)}]{};
      \node[main node] (3) [below = 0.7cm of 2, label=above:{\(W\)}]{};
      \node[main node] (4) [right = 0.7cm of 2, label=above:{\(U_S\)}]{};
      \node[main node] (5) [right = 0.7cm of 3, label=above:{\(U_W\)}]{};
      \node[main node] (6) [below = 0.6cm of 5, label=above:{\(U_Y\)}]{};
      \node[main node] (7) [right = 0.7cm of 5, label=above:{\(W^*\)}]{};  
      \node[main node] (8) [right = 0.7cm of 4, label=above:{\(S^*\)}]{};
      \node[main node] (9) [right = 0.5cm of 8, label=right:{\(X^*\)}]{};
      \node[main node] (10) [below = 1.5cm of 9, label=below:{\(Y^*\)}]{};
      \node[main node] (11) [above = 1cm of 4, label=left:{\(U_x\)}]{};
      \path[every node/.style={font=\sffamily\small}]
        (0) edge node {} (1)
        (3) edge node {} (0)
        (3) edge node {} (1)
        (0) edge node {} (2)
        (9) edge node {} (10)
        (7) edge node {} (9)
        (7) edge node {} (10)
        (9) edge node {} (8);
      \draw [dashed] (4) -- (2);
      \draw [dashed] (4) -- (8);
      \draw [dashed] (5) -- (7);
      \draw [dashed] (5) -- (3);
      \draw [dashed] (6) -- (1);
      \draw [dashed] (6) -- (10);
      \draw [dashed] (11) -- (0);
    \end{tikzpicture}
    \caption{}
    \label{causalg_b}
    \end{subfigure}\hfill
    \begin{subfigure}[b]{0.35\textwidth}
    \centering
    \begin{tikzpicture}[->,>=stealth',node distance=2cm,
      thick,main node/.style={circle,fill,inner sep=1.5pt}]
    
      \node[main node] (0) [label=left:{\(X\)}]{};
      \node[main node] (1) [below =1cm of 0,label=below:\(Y\)] {};
      \node[main node] (2) [above = 1cm of 0, label=above:{\(W\)}]{};
      \node[main node] (3) [above right = 0.7cm of 0, label=above:{\(S\)}]{};
      \node[main node] (4) [right = 1.2cm of 2, label=above:{\(U_W\)}]{};
      \node[main node] (5) [below = 0.7cm of 4, label=above:{\(U_S\)}]{};
      \node[main node] (6) [below = 1.5cm of 4, label=above:{\(U_x\)}]{};
      \node[main node] (7) [below = 2.5cm of 4, label=above:{\(U_Y\)}]{};
      \node[main node] (8) [right = 1.2cm of 4, label=above:{\(W^*\)}]{};
      \node[main node] (9) [below = 1cm of 8, label=above:{\(X^*\)}]{};
      \node[main node] (10) [below =1cm of 9,label=below:\(Y^*\)] {};
      \node[main node] (11) [above left = 0.7cm of 9, label=above:{\(S^*\)}]{};
      \path[every node/.style={font=\sffamily\small}]
        (0) edge node {} (1)
        (2) edge node {} (0)
        (2) edge node {} (3)
        (9) edge node {} (10)
        (8) edge node {} (11);
      \draw [dashed] (4) -- (2);
      \draw [dashed] (4) -- (8);
      \draw [dashed] (5) -- (11);
      \draw [dashed] (6) -- (0);
      \draw [dashed] (5) -- (3);
      \draw [dashed] (7) -- (10);
      \draw [dashed] (7) -- (1);
    \end{tikzpicture}
    \caption{}
    \label{causalg_c}
    \end{subfigure}
}
\caption{Figures (a) and (c) satisfy natural experimental s‑recoverability, whereas in Figure (b), the confounder \(W\) introduces selection bias into the counterfactual variable \(Y^*\).
}
\label{fig2}
\end{figure}
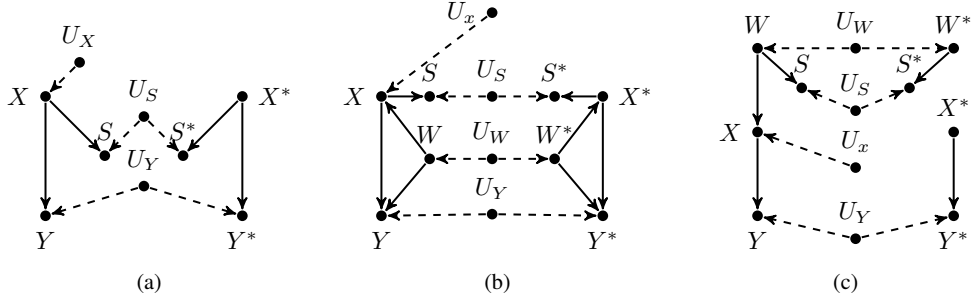
Another key tool employed in our analysis is the twin network introduced by \citet{pearl2009causality}. The twin network is a representation that extends the original causal graph by constructing a parallel counterfactual graph. As shown in Figure~\ref{fig2}, the counterfactual counterparts of the original variables share the same exogenous factors as their factual versions. This construction provides a unified framework to simultaneously reason about factual and counterfactual quantities. In our work, the twin network plays a critical role, as our theoretical analysis and algorithmic developments rely heavily on its structure for properly encoding the relationships between variables and ensuring the validity of d-separation conditions in the counterfactual domain. 

It is worth noting that within the twin network, counterfactual variables are denoted using starred variables; for instance, \(X^*\) and \(Y^*\) represent the counterfactual versions of the treatment and outcome, respectively. However, the counterfactual statement "Variable \(Y\) would have the value \(y\) had \(X\) been \(x\)" is used to be denoted as \(Y_{X=x}=y\), abbreviated as \(y_{x}\)". To prevent confusion, we introduce the starred notation \(P(Y^*_{X^*})\) specifically to denote the experimental distribution in counterfactual logic. 


In an ideal experimental setting free of unmeasured confounding, the distribution of \(P(Y^*_{X^*})\) is identical to the interventional (or experimental) distribution (\citet{pearl2022causal}). Therefore, in the twin network, the distribution \(P(y^*_{x^*})\) can be expressed in the following form:
\[
P(y^*_{x^*}) = P(Y^*_{X^*=x} = y) = P(Y=y | do(X=x)),
\] 
Furthermore, when we refer to the independence between the variable \(S\) and the \(Y^*_{X^*}\), it is equivalent to stating that, within the twin network, the node corresponding to \(S\) is d-separated (\citet{pearl2014probabilistic}) from the node corresponding to \(Y^*_{X^*}\).

\section{Recoverability using counterfactual logic}
In this section, we will systematically discuss how to determine whether the current experimental distribution is affected by selection bias when there is indeed selection bias in the experimental process. Additionally, we will explore how to recover an unbiased experimental distribution using an unbiased distribution provided by partially observable external data when facing a biased experimental distribution.
\subsection{Recoverability without external data}
\label{headings}


\begin{definition}[Natural experimental s‑Recoverability]\label{def1}
Given a causal graph \(G_s\) augmented with a node \(S\) encoding the selection mechanism. The experimental distribution \(Q = P(Y^{*}_{X^*})\) is said to be naturally recoverable in \(G_s\) if, for every experimental distribution \(P(Y^{*}_{X^*})\) compatible with \(G_s\), the following condition holds naturally: \(P(Y^{*}_{X^*} | S = 1) = P(Y^{*}_{X^*}) > 0\).
\end{definition}

It is noteworthy that a causal graph, as an abstract representation of structural causal models, essentially encapsulates a family of structural causal equations that share a consistent causal logic. In practical applications, the true structural causal equations are often difficult to obtain, and their intricate forms and underlying details may incur significant computational complexity while introducing potential biases through additional assumptions. In contrast, employing a nonparametric approach via causal graphs to analyze the impact of selection bias on experimental outcomes enables a more efficient revelation of the global causal structure among variables without being encumbered by the complexities of specific model formulations.

More specifically, by augmenting the original causal graph to construct a twin network and utilizing d-separation as a nonparametric criterion, we can effectively identify those causal graphs that satisfy natural recoverability, or discern the causal pathways and substructures through which selection bias might propagate to the experimental outcomes. In other words, this methodology affords an intuitive and efficient means to analyze the properties of the experimental distribution \(P(Y^*_{X^*})\). 

Consider the Figure \ref{causalg_a}, the data collection process is exclusively associated with the \(X^*\) node. This might suggest that employing a dataset subject to selection bias might significantly perturb the experimental outcome distribution \(P(Y^*_{X^*})\), as evidenced by the inequality \(P(Y^*_{X^*} | S=1) \neq P(Y^*_{X^*})\). However, by constructing a twin network that incorporates shared exogenous variables, it can be demonstrated that the selection bias in Figure \ref{causalg_a} does not affect the true exogenous variable \(U_y\). Consequently, in the corresponding counterfactual domain, the distribution \(P(Y^*_{X^*})\) remains invariant with respect to the selection variable \(S\); that is, the experimental distribution \(P(Y^*_{X^*})\) is determined solely by \(X^*\) and exogenous variable, thereby avoiding the effect of selection bias in this causal graph. More generally,  when experiments are conducted using selection-biased data, the selection mechanism, when regarded as prior information, does not perturb the variables that decide the experimental distribution. Accordingly, the distribution \(P(Y^*_{X^*})\) compatible with Figure~\ref{causalg_a} satisfies natural s-recoverability, obviating the necessity for external data in its recovery.

\begin{figure}[t]
\centering
\resizebox{1.0\textwidth}{!}{%
    \begin{subfigure}[b]{0.35\textwidth}
    \centering
    \begin{tikzpicture}[->,>=stealth',node distance=2cm,
  thick,main node/.style={circle,fill,inner sep=1.5pt}]
      \node[main node] (0) [label=above:{\(S\)}]{};
      \node[main node] (1) [above left = 1cm of 0,label=left:\(X\)]{};
      \node[main node] (2) [below =1.5cm of 1,label=below:\(Y\)] {};
      \node[main node] (3) [above right = 0.6cm of 0, label=above:{\(U_S\)}]{};
      \node[main node] (4) [below = 0.8 cm of 3, label=above:{\(U_Y\)}]{};
      \node[main node] (5) [above right = 0.5 cm of 1, label=above:{\(U_X\)}]{};
      \node[main node] (6) [below right = 0.6cm of 3, label=above:{\(S^*\)}]{};
      \node[main node] (7) [above right = 1cm of 6,label=right:\(X^*\)]{};
      \node[main node] (8) [below = 1.5cm of 7, label=below:{\(Y^*\)}]{};
      \path[every node/.style={font=\sffamily\small}]
        (1) edge node {} (2)
        (2) edge node {} (0)
        (8) edge node {} (6)
        (7) edge node {} (8);
      \draw [dashed] (3) -- (0);
      \draw [dashed] (4) -- (2);
      \draw [dashed] (3) -- (6);
      \draw [dashed] (5) -- (1);
      \draw [dashed] (4) -- (8);
    \end{tikzpicture}
    \caption{}
    \label{causalg_3a}
    \end{subfigure}\hfill
    \begin{subfigure}[b]{0.55\textwidth}
    \centering
    \begin{tikzpicture}[->,>=stealth',node distance=2cm,
      thick,main node/.style={circle,fill,inner sep=1.5pt}]
    
      \node[main node] (0) [label=above:{\(W_1\)}]{};
      \node[main node] (1) [above left = 1cm of 0,label=left:\(X\)]{};
      \node[main node] (2) [below =2.5cm of 1,label=below:\(Y\)] {};
      \node[main node] (6) [right = 3cm of 0, label=above:{\(W_1^*\)}]{};  
      \node[main node] (7) [above right = 1cm of 6,label=right:\(X^*\)]{};
      \node[main node] (8) [below = 2.5cm of 7, label=below:{\(Y^*\)}]{};
      \node[main node] (9) [above right = 1cm of 1, label=left:{\(W_2\)}]{};  
      \node[main node] (10) [above left = 1cm of 7, label=right:{\(W_2^*\)}]{};     
      \node[main node] (11) [above right = 1cm of 0, label=above:{\(W_3\)}]{};
      \node[main node] (12) [above left = 1cm of 6, label=above:{\(W_3^*\)}]{};
      \node[main node] (13) [below = 0.6cm of 11, label=below:{\(S\)}]{};
      \node[main node] (14) [below = 0.6cm of 12, label=below:{\(S^*\)}]{};
      \node[main node] (15) [right = 0.6cm of 11, label=above:{\(U_{W_3}\)}]{};
      \node[main node] (16) [above = 0.65cm of 15, label=above:{\(U_{W_2}\)}]{};
      \node[main node] (17) [below = 0.65cm of 15, label=above:{\(U_S\)}]{};
      \node[main node] (18) [below = 1.7cm of 15, label=below:{\(U_{W_1}\)}]{};
      \node[main node] (19) [below = 2.7cm of 15, label=below:{\(U_{Y}\)}]{};
      \node[main node] (20) [above left = 0.6cm of 1, label=above:{\(U_{X}\)}]{};
      \path[every node/.style={font=\sffamily\small}]
        (1) edge node {} (2)
        (0) edge node {} (1)
        (0) edge node {} (2)    
        (7) edge node {} (8)
        (1) edge node {} (9)
        (7) edge node {} (10)    
        (11) edge node {} (9)
        (12) edge node {} (10)
        (6) edge node {} (8)
        (11) edge node {} (13)
        (12) edge node {} (14) 
        ;
      \draw [dashed] (15) -- (11);
      \draw [dashed] (15) -- (12);
      \draw [dashed] (16) -- (9);
      \draw [dashed] (16) -- (10);
      \draw [dashed] (17) -- (13);
      \draw [dashed] (17) -- (14); 
      \draw [dashed] (18) -- (0);
      \draw [dashed] (18) -- (6); 
      \draw [dashed] (19) -- (2); 
      \draw [dashed] (19) -- (8); 
      \draw [dashed] (20) -- (1); 
    \end{tikzpicture}
    \caption{}
    \label{causalg_3b}
    \end{subfigure}\hfill
    \begin{subfigure}[b]{0.55\textwidth}
    \centering
    \begin{tikzpicture}[->,>=stealth',node distance=2cm,
      thick,main node/.style={circle,fill,inner sep=1.5pt}]

      \node[main node] (0) [label=below left:{\(X\)}]{};
      \node[main node] (1) [below =2.5cm of 0,label=left:\(Y\)] {};
    
      \node[main node] (3) [below left =1cm of 1,label=left:\(W_2\)] {};
      \node[main node] (4) [above right =1.5cm of 0,label=left:\(S\)] {};
      \node[main node] (5) [below =1.5cm of 4,label=left:\(W_3\)] {};
      \node[main node] (6) [below =1cm of 5,label=left:\(W_4\)] {};  
      \node[main node] (7) [right =4cm of 0, label=above:{\(X^*\)}]{};
      \node[main node] (8) [below =2.5cm of 7,label=right:\(Y^*\)] {};
      \node[main node] (9) [above left =1.5cm of 7,label=right:\(S^*\)] {};
      \node[main node] (10) [below =1.5cm of 9,label=right:\(W_3^*\)] {};
      \node[main node] (11) [below =1cm of 10,label=right:\(W_4^*\)] {};
    
      \node[main node] (13) [below right =1cm of 8,label=right:\(W_2^*\)] {};
      \node[main node] (14) [right =1.95cm of 1,label=below:\(U_Y\)] {};  
      \node[main node] (15) [above =0.7cm of 14,label=below:\(U_{W_4}\)] {}; 
      \node[main node] (16) [above =1.7cm of 14,label=below:\(U_{W_3}\)] {};
      \node[main node] (17) [below =0.65cm of 14,label=below:\(U_{W_2}\)] {};
      \node[main node] (18) [above =3cm of 14,label=below:\(U_{S}\)] {};
      \node[main node] (19) [left =1.9cm of 4,label=left:\(W_1\)] {};
      \node[main node] (20) [right =1.8cm of 9,label=right:\(W_1^*\)] {};
      \node[main node] (21) [above =4.5cm of 14,label=below:\(U_{W_1}\)] {};
      \node[main node] (22) [above =0.55cm of 0,label=above:\(U_{X}\)] {};  
    
      \path[every node/.style={font=\sffamily\small}]
        (0) edge node {} (1)
        (7) edge node {} (8)
        (19) edge node {} (0)    
        (19) edge node {} (3)
        (20) edge node {} (13)
        (3) edge node {} (1)
        (13) edge node {} (8)
        (3) edge node {} (1)
        (13) edge node {} (8)
        (0) edge node {} (4)
        (7) edge node {} (9)    
        (5) edge node {} (4)
        (10) edge node {} (9)
        (6) edge node {} (5)
        (11) edge node {} (10)
        (6) edge node {} (1)
        (11) edge node {} (8);
      \draw [dashed] (18) -- (9);
      \draw [dashed] (18) -- (4);
      \draw [dashed] (14) -- (8);
      \draw [dashed] (14) -- (1);
      \draw [dashed] (15) -- (11);
      \draw [dashed] (15) -- (6); 
      \draw [dashed] (16) -- (10);
      \draw [dashed] (16) -- (5); 
      \draw [dashed] (17) -- (3); 
      \draw [dashed] (17) -- (13); 
      \draw [dashed] (21) -- (19); 
      \draw [dashed] (21) -- (20); 
      \draw [dashed] (22) -- (0);   
    \end{tikzpicture}
    \caption{}
    \label{causalg_3c}
    \end{subfigure}
}
\caption{Figures (b) satisfies natural experimental s‑recoverability, whereas in Figure (a) and (c), Selection bias is introduced into the counterfactual variable \(Y^*\).
}
\label{fig3}
\end{figure}
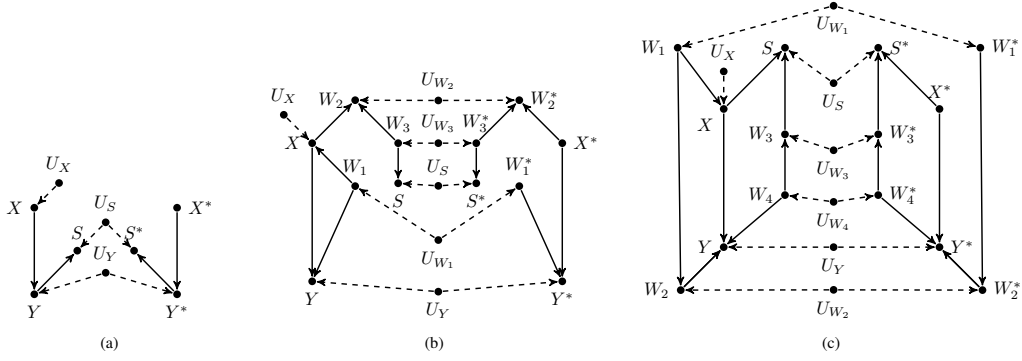

\begin{lemma}{Confounder Irrelevance for Natural experimental s‑Recoverability}\label{lemma1}

In a causal DAG \(G_s\), the presence of a node \(W\) that confounds \(X\) and \(Y\) (i.e., with edges \(W \to X\) and \(W \to Y\)) neither necessarily prevents the natural s–recoverability of \(P(Y^*_{X^*})\), nor does the absence of any such \(W\) necessarily ensure its natural s–recoverability.

\end{lemma}

\emph{Proof:} Consider Figure~\ref{causalg_c}, in which a confounder \( W \) exists between nodes \( X \) and \( S \). This suggests the possibility that selection bias may propagate to the distribution \( P(Y^*_{X^*}) \) through node \( W \) and its associated exogenous variable \( U_W \). However, under intervention on node \( X^* \) in the counterfactual scenario, no active path exists between nodes \( S \) and \( Y^*_{X^*} \). Consequently, the equality \( P(Y^*_{X^*} | S=1) = P(Y^*_{X^*}) \) remains valid. Therefore, Figure~\ref{causalg_3b} satisfies natural counterfactual s-recoverability while still containing a confounder in the causal graph structure.

Consider Figure~\ref{causalg_3a}, in which no confounders are present. Nevertheless, by constructing a twin network via shared exogenous variables, it becomes evident that selection bias can directly influence the experimental distribution \(P(Y^*_{X^*})\) through the \(Y\) node. Consequently, the presence of confounders is neither necessary nor sufficient for natural s-recoverability.


\begin{theorem}\label{thm1}
The distribution \(P(Y^*_{X^*})\) is naturally s-recoverable from \(G_s\) if \((S\perp\!\!\!\perp Y^*_{X^*})\).
\end{theorem}

\emph{Proof:} It is obvious that if \(X\) d-separates \(S\) from \(Y^*\) in \(G_s\), \(P (Y^*_{X*})\) is nature counterfactual s-recoverable.

Theorem \ref{thm1} provides an efficient and straightforward criterion for verifying whether a distribution satisfies natural s-recoverability. Specifically, it indicates that no external data are necessary for recovering the distribution \( P(Y^*_{X^*}) \). The procedure involves constructing a twin network with shared exogenous variables and examining whether the nodes \( S \) and \( Y^*_{X^*} \) are d-separated given the empty set. If the d-separation condition \( S \perp\!\!\!\perp Y^*_{X^*} | \emptyset \) holds in the twin network, then the distribution \( P(Y^*_{X^*}) \) satisfies natural \( s \)-recoverability.

Although, according to Lemma~\ref{lemma1}, the absence of confounders is neither a sufficient nor a necessary condition for natural counterfactual s-recoverability. However, treating confounder nodes as potential mediators for transmission of selection bias is reasonable, and confounder is highly likely to transmit selection bias to the experimental distribution. Consider Figure~\ref{causalg_b}. By constructing a twin network via shared exogenous nodes, it becomes evident that the selection node \(S\) exerts influence on the experimental distribution \(P(Y^*_{X^*})\) through a spurious pathway mediated by the confounding variable \(W\) and its counterfactual counterpart \(W^{*}\). Specifically, when a confounder \(W\) induces a spurious association between nodes \(X\) and \(Y\), the selection node \(S\) will influence the experimental distribution \(P(Y^*_{X^*})\) via the shared exogenous variable \(U_w\). Consequently, this mechanism leads to the activation of selection bias, violating the condition of natural s-recoverability for \(P(Y^*_{X^*})\).

\subsection{Recoverability with external data}
When the experimental distribution \( P(Y^*_{X^*}) \), compatible with a given causal graph \( G_s \), fails to satisfy natural experimental  s-recoverability, does this imply permanent impossibility in recovering \( P(Y^*_{X^*}) \)? Not necessarily. If we have access to external unbiased data, such as the distribution \( P(\text{Gender}) \) easily obtainable from population census records, there exists an opportunity for recovering the experimental distribution \( P(Y^*_{X^*}) \). In this section, we systematically analyze how to determine, from a known causal graph, the precise types of external unbiased data required for restoring the experimental distribution \( P(Y^*_{X^*}) \). Furthermore, we propose a concrete algorithm for identifying the set of external data variables necessary to recover the experimental distribution \( P(Y^*_{X^*}) \).



Despite the elegant property of natural counterfactual s-recoverability, that is, the fact that \(P(Y^*_{X^*})=P(Y^*_{X^*}| S=1)\) and no external data are required, in practice we often encounter scenarios where external data are necessary to recover \(P(Y^*_{X^*})\) from \(P(Y^*_{X^*}| S=1)\). 

Consider Figure~\ref{causalg_3c}. Suppose our ultimate goal is to recover the experimental distribution \(P(Y^*_{X^*})\) encoded in Figure~\ref{causalg_3c}. By constructing a twin network via shared exogenous variables, Theorem~\ref{thm1} immediately reveals that the distribution \(P(Y^*_{X^*})\) in Figure~\ref{causalg_3c} does not satisfy natural counterfactual s-recoverability. Consequently, the only viable strategy is to incorporate external data to recover an unbiased experimental distribution. In particular, the set \(\{W_1,W_3\}\) d-separates the selection node \(S\) from the counterfactual node \(Y^*_{X^*}\), implying that external measurements of \(W=\{W_1,W_3\}\) are sufficient for recovery. Therefore, the target experimental distribution can be expressed as
\[
\begin{aligned}
P(Y^*_{X^*}) 
&= \sum_{w_1,w_3} P\bigl(Y^*_{X^*} | w_1, w_3\bigr)\,P(w_1,w_3) \\
&= \sum_{w_1,w_3} P\bigl(Y^*_{X^*} | w_1, w_3, S=1\bigr)\,P(w_1,w_3).
\end{aligned}
\]
The validity of above equation arises from the conditional independence relation \(Y^*_{X^*} \perp\!\!\!\perp S | \{W_1, W_3\}\) within the twin network constructed via shared exogenous variables. Consequently, the experimental distribution \(P(Y^*_{X^*})\) can be explicitly decomposed into two components: the first being the biased experimental data distribution \(P(Y^*_{X^*}| W_1,W_3,S=1)\), and the second representing unbiased observational data \(P(W_1,W_3)\). Hence, in Figure~\ref{causalg_3c}, the inclusion of unbiased observational data allows us to recover the unbiased experimental distribution from biased experimental data. More abstractly, this recovery procedure can be viewed as a correction mechanism, in which unbiased observational distributions are employed to adjust the biased experimental distribution.


\begin{definition}{General experimental s-recoverability}\label{def:general-s-recov}
Let \(G_s\) be a causal graph augmented with a selection node \(S\), and let \(V\) denote the set of observed variables. Suppose \(M, W \subseteq V\), where \(M\) (with distribution \(P(M | S=1)\)) represents the biased experimental data and \(W\) (with distribution \(P(W)\)) represents the unbiased data, allowing \(W = \emptyset\). We say that the experimental distribution \(P(Y^*_{X^*})\) is generally s-recoverable in \(G_s\) if, for any two distributions \(P_1\) and \(P_2\) that are compatible with \(G_s\) and satisfy \( P_1(M | S=1) \;=\; P_2(M | S=1) > 0\quad \text{and} \quad P_1(W) \;=\; P_2(W)> 0, \) it follows that \( P_1(Y^*_{X^*}) \;=\; P_2(Y^*_{X^*}). \)
\end{definition}

 The example in Figure~\ref{causalg_3c} illustrates that one can attempt to identify a set of variables measurable at the population level in order to obtain an unbiased distribution. Under ideal conditions, such an unbiased observational distribution can be leveraged to recover the biased experimental distribution \(P(Y^*_{X^*})\). However, in practice it is unrealistic to assume that every node is measurable at the population level. For instance, if \(P(W_3)\) is unobservable, is there a method to determine whether an equivalent set exists within the current causal graph that ensures the s-recoverability of \(P(Y^*_{X^*})\)?

\begin{algorithm}[H]
\caption{General experimental s-recoverability of $P(Y^*_{x^*})$}\label{alg:twin}
\begin{algorithmic}[1]
\REQUIRE External unbiased variable set $W$
\STATE \textbf{Twin Network Construction:} Create a twin network by sharing exogenous variables and remove all edges entering the counterfactual node $X'$.
\IF{$Y^*_{x^*} \perp\!\!\!\perp S | \emptyset$}
    \STATE \textbf{return} $P(Y^*_{x^*})$ is naturally experimental s-recoverable.
\ENDIF
\FOR{each set $Z \in M$}
    \IF{$Y^*_{x^*} \perp\!\!\!\perp S | Z$

$$        
P(Y^*_{x^*}) = P(Y^*_{x^*} | Z, S=1) \, P(Z).
$$}

        \IF{$Z \in W$}
            \STATE \textbf{return} $P(Y^*_{x^*})$ is experimental s-recoverable.
        \ELSE
            \STATE Call RC(\(Z, \emptyset\)) (See Appendix for Algorithm RC)
                    \IF{RC(\(Z, \emptyset\)) is True}
                    \STATE \textbf{return} $P(Y^*_{x^*})$ is experimental s-recoverable.
                    \ELSE
                    \STATE \textbf{return} \texttt{FAILURE}.
                    \ENDIF
        \ENDIF
    \ENDIF
\ENDFOR
\STATE \textbf{return} \texttt{FAILURE}.
\end{algorithmic}
\end{algorithm}

According to Algorithm~\ref{alg:twin}, it is straightforward to deduce that in Figure~\ref{causalg_3c} the set \(\{W_1, W_4\}\) is also a valid candidate for ensuring that \(P(Y^*_{X^*})\) satisfies s-recoverability. Thus, if \(W_3\) is unobservable at the population level, \(W_4\) may serve as an alternative, preserving the possibility that \(P(Y^*_{X^*})\) remains s-recoverable. Moreover, Algorithm~\ref{alg:twin} provides experimenters with a flexible recovery strategy, allowing them to select the admissible set that is most advantageous, convenient, and cost-effective for achieving unbiased s-recovery of \(P(Y^*_{X^*})\).

\begin{theorem}\label{thm2}
Let \(G_s\) be a causal graph augmented with a selection node \(S\), and let \(V\) denote the set of all variables. Suppose there exists a subset \(Z \subseteq V\) that is measured in both the biased experiment and at the population level, and that \(Y^*_{X^*} \perp\!\!\!\perp S | Z\). Then, the experimental distribution is s-recoverable: \(P(Y^*_{X^*}) \;=\; \sum_{z} P(Y^*_{X^*} | Z, S=1)\, P(Z).\)
\end{theorem}

Theorem~\ref{thm2} precisely characterizes how biased experimental distributions can be systematically integrated with external unbiased distributions, yielding a straightforward yet generalizable recovery method. This theorem not only establishes a theoretically sound and directly implementable criterion for recovering experimental distributions but also provides a practical roadmap for empirical research design and data analysis.

\begin{lemma}
If experimental distribution \(P(Y^*_{X^*})\) in \(G_s\) is not s-recoverable, then \(P(Y^*_{X^*})\) is not s-recoverable in the graph \(G_s^{'}\) resulting from adding a single edge to \(G_s\).
\end{lemma}

This illustrates that when it is determined that a graph structure does not satisfy counterfactual s-recoverability, simply by adding structural information to this graph will not help the graph obtain counterfactual s-recoverability. Therefore, when we exclude a graph from satisfying counterfactual s-recoverability, we also exclude a class of graphs derived by adding edges to this graph, This provides us with a convenient condition for judging complex graphs. Once we find that a subgraph of the complex graph violates counterfactual s-recoverability, it is equivalent to the complex graph violating counterfactual s-recoverability, because the complex graph can be regarded as recursively adding an edge to the subgraph.


\section{Experiments}
\subsection{Discrete example}\label{exp_dis}
We consider a scenario involving the assessment of a novel medicine aimed at treating a specific type of pneumonia and there are not enough clinical observational data about the novel medicine available. Recovery from this disease is known to depend jointly on the administration of the novel treatment, the presence of potential comorbidities, and disease severity. Researchers aim to determine whether the new treatment is generally superior in effectiveness compared to a standard generic drug. To study this question, real-world patients are recruited into an experimental group with probabilities dependent explicitly on their severity levels: severely ill patients have a 70\% probability of inclusion, whereas mildly ill patients have only a 30\% probability. Consequently, this differential selection process systematically induces selection bias, posing significant methodological challenges for the unbiased estimation of treatment efficacy.

\noindent\textbf{Notation:} Let 
$X\in\{0,1\}$ be the treatment indicator (1=novel drug, 0=standard);\  
$W\sim\mathrm{Bern}(0.5)$ the comorbidity marker (1=present, 0=absense);\  
$Z\sim\mathrm{Bern}(0.5)$ the disease severity (1=severe, 0=mild);\  
$S\in\{0,1\}$ the selection indicator with 
$P(S=1| Z{=}1)=0.7,\,P(S=1| Z{=}0)=0.3$;\  
and $Y\in\{0,1\}$ the recovery outcome (1=recover, 0=failure).

  

Table~\ref{tab:probabilities} provides the ideal distribution underlying our experimental setup, while Table~\ref{tab:counts} summarizes the biased experimental subgroup information. In our experiment, patients were randomly assigned with equal probability to either the standard therapy or the novel drug.

\begin{table}[ht]
\caption{The ideal distribution information and biased dataset for the experiment}
\centering
\small
\setlength{\tabcolsep}{4pt}

\begin{minipage}{0.45\textwidth}
\centering
\begin{tabular}{@{}ccrr@{}}
\toprule
$W$ & $Z$ & $P(Y=1| X=0,w,z)$ & $P(Y=1| X=1,w,z)$ \\
\midrule
0 & 0 & 0.90 & 0.95 \\
0 & 1 & 0.50 & 0.80 \\
1 & 0 & 0.70 & 0.90 \\
1 & 1 & 0.30 & 0.60 \\
\bottomrule
\end{tabular}
\subcaption{Theoretical recovery probabilities by subgroup}
\label{tab:probabilities}
\end{minipage}
\hfill
\begin{minipage}{0.45\textwidth}
\centering
\begin{tabular}{@{}cccrr@{}}
\toprule
$X$ & $W$ & $Z$ & Not Recovered & Recovered \\
\midrule
0 & 0 & 0 &  12 & 141 \\
0 & 0 & 1 & 174 & 180 \\
0 & 1 & 0 &  42 & 100 \\
0 & 1 & 1 & 245 & 110 \\
1 & 0 & 0 &   8 & 158 \\
1 & 0 & 1 &  73 & 266 \\
1 & 1 & 0 &  10 & 146 \\
1 & 1 & 1 & 146 & 218 \\
\bottomrule
\end{tabular}
\subcaption{Recovery information in experiment group with selection bias}
\label{tab:counts}
\end{minipage}
\label{tab:side_by_side}
\end{table}

To obtain the theoretical distribution \(P(Y^*_{x^*})\), we apply the back‐door adjustment over the risk marker \(W\) (See Appendix for detailed calculation):

\[
\begin{aligned}
P(Y^*_{x^*})
&= \sum_{w\in\{0,1\}} P\bigl(Y | do(X = x), W = w\bigr)\,P(W = w) \\
&\implies
\begin{cases}
P(Y^*_{x=1} = 1)
= 0.8125;  P(Y^*_{x=1} = 0) = 0.1875\\[6pt]
P(Y^*_{x=0} = 1)
= 0.60;  P(Y^*_{x=0} = 0) = 0.40\\
\end{cases}
\end{aligned}
\]

By Theorem~\ref{thm2} . We only need external distribution for Z to restore the biased experimental distribution to an unbiased one. From the open external source dataset, researchers know that \(Z\sim\mathrm{Bernoulli}(0.5)\).
\[
\begin{aligned}
P_{rec}(Y^*_{x^*})
&= \sum_{z} P\bigl(Y^*_{x^*} | Z=z,\,S=1\bigr)\,P(Z=z)\\
&\implies
\begin{cases}
P(Y_{x=1} = 1)
\approx 0.816;  P(Y_{x=1} = 0)\approx 1-0.816=0.184 \\[6pt]
P(Y_{x=0} = 1)
\approx 0.613; P(Y_{x=0} = 0)\approx 1-0.613 = 0.387\\
\end{cases}
\end{aligned}
\]

\begin{table}[h]
\centering
\caption{Biased experimental distribution and relative errors}
\label{tab:re_and_errors}
\resizebox{0.8\textwidth}{!}{%
\begin{tabular}{@{}l
                cc
                cc@{}}
\toprule
Treatment & 
\multicolumn{2}{c}{\(P(Y^*_{x^*}| S=1)\)} & 
\multicolumn{2}{c}{Relative Error} \\
\cmidrule(lr){2-3} \cmidrule(lr){4-5}
\(X\) & \(P(Y^*_{X^*}=1 
| S=1)\) & \(P(Y^*_{X^*}=0 | S=1)\) & \(\mathrm{RE}_{\mathrm{bias}}\) & \(\mathrm{RE}_{\mathrm{rec}}\) \\
\midrule
Standard (\(X=0\)) & 
\(\frac{531}{531+473}\approx0.529\) & \(1-0.529=0.471\) & \(-11.8\%\) & \(+2.2\%\) \\[4pt]
Novel    (\(X=1\)) & 
\(\frac{788}{788+237}\approx0.768\) & \(1-0.768=0.232\) & \(-5.5\%\) & \(+0.4\%\) \\
\bottomrule
\end{tabular}
}
\end{table}

The researchers calculated the relative errors based on the recovery of the experimental distribution \(P_{rec}(Y^*_{x^*})\)  and the biased experimental distribution, respectively. For detailed data, see Table \ref{tab:re_and_errors}. We observe that the relative error of the recovered experimental distribution (\(\mathrm{RE_{rec}}\)) is substantially smaller than that of the biased observed distribution (\(\mathrm{RE_{bias}}\)). Specifically, the relative error for the standard treatment improves notably from \(-11.8\%\) to \(+2.2\%\), while for the novel treatment it improves from \(-5.5\%\) to merely \(+0.4\%\). This simulated experiment thus demonstrates the practical effectiveness of leveraging external distribution information to correct for selection bias, validating the applicability of our theoretical framework in realistic settings.

\subsection{Continuous example}\label{continuous_example}


In this study, we simulate a clinical trial designed to evaluate a novel therapy for a specific pulmonary condition without enough observational data. Participants are recruited based on their baseline inflammatory biomarker levels, denoted by \(Z\). During enrollment, researchers preferentially select units with higher \(Z\) values, thereby introducing systematic selection bias. Once enrolled, treatment assignment \(X\) (novel drug:X=1 vs. standard care:X=0) is randomized via a Bernoulli draw. We formalize this data‑generating process with the following structural causal model (SCM), which serves as our ground‑truth SCM  for subsequent simulation studies. (See Appendix for corresponding causal graph) 

\begin{table}[h]
  \centering
  \caption{Summary of our SCM variables.}
  \small
  \begin{tabular}{@{}l@{\quad}l@{\quad}l@{}}
    \toprule
    Symbol & Meaning & Generation \\
    \midrule
    \(X\) & Treatment
        & \(X=\mathbf{1}\{\gamma_{WX}W + U_X > 0\};\ U_X\sim\mathrm{Uniform}(0,1)\) \\
    \(W\) & Latent health (e.g.\ prior lung function) 
        & \(W=U_W\); \(U_W\sim\mathcal{N}(0,1)\) \\
    \(Z\) & Baseline inflammation severity 
        & \(Z=U_Z\); \(U_Z\sim\mathcal{N}(0,1)\) \\
    \(Y\) & Observed change in inflammation 
        & \(Y=\alpha X + \beta Z + \gamma_{WY}W + U_Y\); \(U_Y\sim\mathcal{N}(0,\sigma_Y^2)\) \\
    \(S\) & Selection indicator 
        & \(S=\mathbf{1}\{\gamma_Z Z + U_S > c\}\); \(U_S\sim\mathcal{N}(0,\sigma_S^2)\) \\
    \bottomrule
  \end{tabular}
  \label{tab:scm_summary}
\end{table}

To better reflect realistic constraints, we assume that investigators can collect biased experimental cohorts of sizes \(n\in\{100,200,500,1000,2000,4000\}\).  For each \(n\), we draw 50 independent samples (using distinct random seeds) from the full synthetic dataset, computing and recording the average recovered experimental distribution \(\hat P_{\mathrm{rec}}(Y^*_{x^*})\), its average error relative to the ground truth, and the average biased experimental distribution \(\hat P_{\mathrm{bias}}(Y^*_{x^*})\).  Crucially, from the researchers’ perspective only the biased experimental data and a limited set of external unbiased observational measurements are available; all performance metrics are evaluated under this information regime.

According to Theorem \ref{thm2}, \(P(y^*_{x^*}) = P(y^*_{x^*}|Z,S=1)P(Z)\). Furthermore, since intervening on \(X\) does not effect the distribution of \(S=1\) (as the distribution of \(S=1\) depends solely on \(Z\), it follows that\(P(y^*_{x^*}| Z,S=1)=P(y| do(x),Z,S=1)\) , this conditional distribution can be directly estimated from the biased experimental data collected by researchers using kernel density estimation (KDE) (See Appendix for KDE graph).
\begin{figure}[h]        
  \centering
  \includegraphics[width=0.8\linewidth]{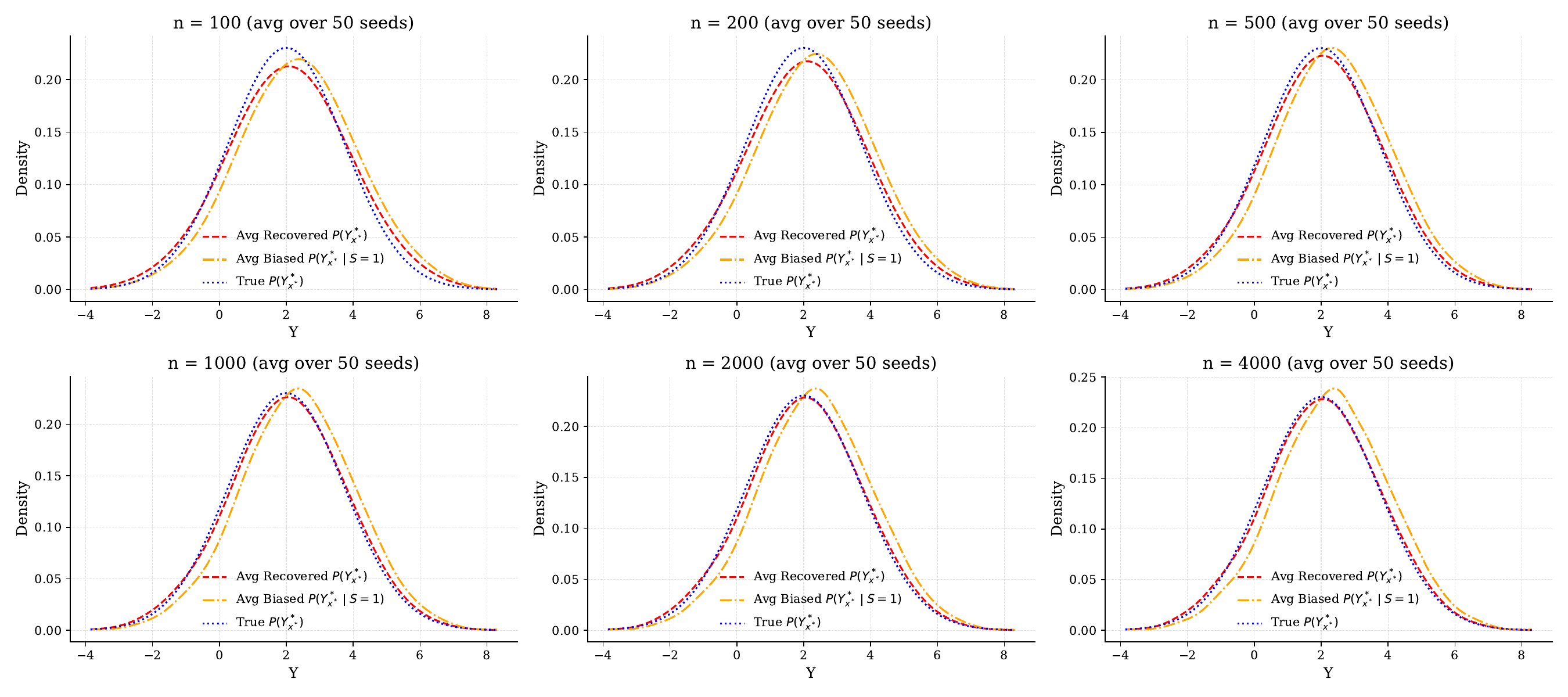}
  \caption{Density comparison of average recovered \(\overline{P}(Y^*_{x^*})\), average conditional \(\overline{P}(Y^*_{x^*}| S=1)\), and theoretical \(P(Y^*_{x^*})\) for sample sizes \(n\in\{100,200,500,1000,2000,4000\}\).}
  \label{fig:recovergraph}
\end{figure}

\begin{figure}[h]        
  \centering
  \includegraphics[width=1.0\linewidth]{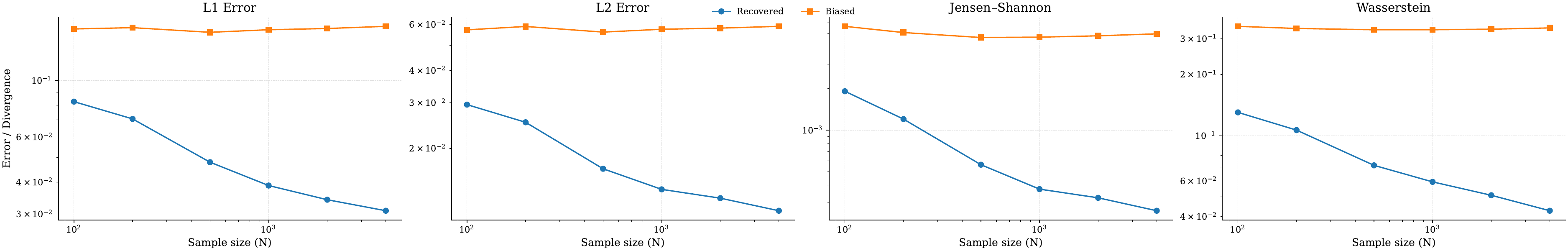}
  \caption{Comparison of averaged error metrics between the recovered experimental distribution and the biased follow‑up distribution across sample sizes \(n\). Figures (a)–(d) display, respectively, (a) L1 error, (b) L2 error, (c) Jensen–Shannon divergence, and (d) Wasserstein distance, averaged over 50 random seeds. }
  \label{fig:error_new}
\end{figure}

\begin{table}[h]
\centering
\caption{Error metrics comparing recovered and biased distributions.}
\label{tab:error_metrics_4_2}
\resizebox{0.8\textwidth}{!}{%
    \begin{tabular}{rrrrrrrrr}
    \toprule
    N & $L1_{rec}$ & $L1_{bias}$ & $L2_{rec}$ & $L2_{bias}$ & $JS_{rec}$ & $JS_{bias}$ & $Wass_{rec}$ & $Wass_{bias}$ \\
    \midrule
    100 & 0.0826 & 0.1590 & 0.0295 & 0.0573 & 0.0019 & 0.0056 & 0.1302 & 0.3446 \\
    200 & 0.0707 & 0.1608 & 0.0252 & 0.0590 & 0.0012 & 0.0051 & 0.1065 & 0.3364 \\
    500 & 0.0479 & 0.1542 & 0.0167 & 0.0562 & 0.0006 & 0.0047 & 0.0715 & 0.3315 \\
    1000 & 0.0388 & 0.1578 & 0.0139 & 0.0576 & 0.0004 & 0.0047 & 0.0593 & 0.3316 \\
    2000 & 0.0341 & 0.1596 & 0.0129 & 0.0582 & 0.0003 & 0.0048 & 0.0510 & 0.3337 \\
    4000 & 0.0309 & 0.1628 & 0.0115 & 0.0591 & 0.0003 & 0.0050 & 0.0428 & 0.3384 \\
    \bottomrule
    \end{tabular}
}
\end{table}

From Figure \ref{fig:recovergraph}, the average recovered distribution steadily approaches the theoretical true distribution with increasing sample size, while the biased distribution remains consistently far from the true distribution, illustrating our method's capability to accurately reconstruct experimental distributions under biased sampling. Furthermore, Table \ref{tab:error_metrics_4_2} and Figure \ref{fig:error_new} show rapid decreases in error metrics for the recovered distribution as sample sizes grow, indicating clear convergence and consistency; by contrast, errors in the biased distribution remain stable at high levels without signs of convergence. These results confirm the efficacy and statistical consistency of our proposed nonparametric approach in addressing selection bias.

\section{Conclusion}
In this work, leveraging Pearl's twin-network construction, we provide a clear, rigorous framework for recovering experimental distributions under systematic selection bias. we first introduce Theorem~\ref{thm1} to determine which selection bias leaves the experimental distribution \(P(Y_x)\) invariant. We then introduce Theorem~\ref{thm2} specifying precisely when unbiased experimental distributions can be reconstructed using biased subgroup experimental data combined with external unbiased observational distributions. Additionally, our proposed algorithm systematically identifies the valid set of external unbiased variables required for accurate recovery.

Extensive simulation studies demonstrate the stability and efficiency of our approach across varying sample sizes and random seeds. Compared to uncorrected biased estimates, our recovered experimental densities converge rapidly to the ground truth, significantly reducing multiple error metrics. We believe this unified framework provides a robust theoretical and practical foundation for reasoning in complex, non-randomized sampling environments.

Although our approach systematically recovers unbiased experimental distributions, it still relies on precise knowledge of external unbiased distributions and accurate conditional density estimation in high-dimensional or small-sample scenarios. Future research will focus on relaxing these strong identification requirements by considering weaker assumptions, such as partial or bounded knowledge of conditional distributions or identifiable bounds, to enhance practical applicability and robustness.

\bibliographystyle{plainnat}
\bibliography{reference}



\clearpage
\appendix
\section{Appendix}
\subsection{Background}
Our entire research is built on an understanding of the third level of causal inference: counterfactuals. Pearl (\citet{pearl2000models}) introduces the three‐level causal hierarchy : association, intervention, and counterfactual, commonly known as the “Ladder of Causation”. Therefore, we will introduce the background of causal inference to understand the observational and interventional distribution and experimental distribution, which are frequently mentioned in the paper, from a causal inference perspective.

\begin{definition}[d-separation (\citet{pearl1988probabilistic})] Let \(X\), \(Y\), and \(Z\) be three disjoint subsets of nodes in a DAG \(D\).  Then \(Z\) is said to \emph{d‑separate} \(X\) from \(Y\), denoted
\(I(X, Z, Y)_D,\) \emph{if and only if} there is \emph{no} undirected path from a node in \(X\) to a node in \(Y\) along which all of the following hold:
\begin{enumerate}
  \item Every node on the path with two arrowheads meeting (“collider”) either is in \(Z\) or has a descendant in \(Z\).
  \item Every other node on the path is outside \(Z\).
\end{enumerate}
if and only if \(Z\) blocks every path from a node in \(X\) to a node in \(Y\) and is denoted by
\(
  Y \,\perp\!\!\!\perp\, X | Z.
\)
\end{definition}
\begin{theorem}[Soundness \& Completeness of d‑separation (\citet{pearl2014probabilistic}, \cite{geiger1990identifying}, \citet{verma2022equivalence})]
Let \(G\) be a DAG and \(P\) a joint distribution over its nodes.  If \(P\)
satisfies the global Markov property w.r.t.\ \(G\) and the faithfulness
assumption, then for any disjoint node sets \(X,Y,Z\subseteq V(G)\),

\[
  X \perp_d Y | Z
  \quad\Longleftrightarrow\quad
  X \perp\!\!\!\perp Y | Z.
\]

\end{theorem}

\begin{definition}[do‑Operator (\citet{pearl2009causality})]
Let \(G\) be a causal DAG over variables \(\mathbf V\).  For any subset \(\mathbf X\subseteq\mathbf V\) and values \(\mathbf x\), the intervention \(do(\mathbf X=\mathbf x)\) is defined by:
\begin{enumerate}
  \item Remove all incoming edges into each node in \(\mathbf X\) to obtain the mutilated graph \(G_{do(\mathbf X)}\).
  \item Fix each \(X\in\mathbf X\) to the value \(x\), while all other variables remain governed by their original structural equations.
\end{enumerate}
The resulting interventional (or experimental) distribution is
\[
  P\bigl(Y | do(\mathbf X=\mathbf x),\,\mathbf Z=\mathbf z\bigr)
  \;=\;
  P_{G_{do(\mathbf X)}}\bigl(Y | \mathbf Z=\mathbf z,\;\mathbf X=\mathbf x\bigr),
\]
which generally differs from the observational conditional \(P(Y| \mathbf X=\mathbf x,\mathbf Z=\mathbf z)\).
\end{definition}

\begin{definition}[Counterfactuals {\citet{pearl2000models}}]
Given a structural causal model \(M\) and observed evidence \(e\), a counterfactual query  
\[
  Y_{x'}(u) = y
\]
is read as “had we set \(X\) to \(x'\) in the unique background context \(u\) consistent with \(e\), \(Y\) would (or would not) take value \(y\).”  Formally:
\begin{enumerate}
  \item Identify the unique exogenous assignment \(u\) satisfying the evidence \(e\).
  \item Modify the model \(M\) by replacing the structural equations of each \(X\in\mathbf X\) with the constant \(x'\), yielding the mutilated model \(M_{x'}\).
  \item Evaluate the sentence \(\bigl(Y(u) = y\bigr)\) in \(M_{x'}\).
\end{enumerate}
Pearl defines this as “the ultimate level of causal hierarchy” and denotes such queries as  
\[
  P\bigl(Y_{x'} = y | e\bigr).
\]
\end{definition}

\begin{definition}[s-recoverability(\citet{bareinboim2022recovering})]
Given a causal graph $G_s$ augmented with a node $S$ encoding the selection mechanism \citet{bareinboim2012controlling}, the distribution $Q = P(y | x)$ is said to be \emph{s-recoverable} from selection-biased data in $G_s$ if the assumptions embedded in the causal model render $Q$ expressible in terms of the distribution under selection bias $P(\textbf{v} | S = 1)$. Formally, for any two probability distributions $P_1$ and $P_2$ that are compatible with $G_s$, if
\[
P_1(\textbf{v} | S = 1) \;=\; P_2(\textbf{v} | S = 1) \;>\; 0,
\]
then
\[
P_1(y | x) \;=\; P_2(y | x).
\]
\end{definition}

\begin{definition}[s‑Recoverability with external data (\citet{bareinboim2022recovering})]
Given a causal graph \(G_S\) augmented with a node \(S\), the distribution \(Q = P(y | x)\) is said to be \emph{s‑recoverable} from selection bias in \(G_S\) with external information over \(\mathbf{T} \subseteq \mathbf{V}\) and selection‑biased data over \(\mathbf{M} \subseteq \mathbf{V}\) (for short, s‑recoverable) if the assumptions embedded in the causal model render \(Q\) expressible in terms of \(P(m | S = 1)\) and \(P(t)\), both positive.  Formally, for every two probability distributions \(P_1\) and \(P_2\) compatible with \(G_S\), if they agree on the available distributions,
\[
  P_1(m | S = 1) \;=\; P_2(m | S = 1) \;>\; 0,
  \quad
  P_1(t) \;=\; P_2(t) \;>\; 0,
\]
then they must agree on the query distribution,
\[
  P_1(y | x) \;=\; P_2(y | x).
\]
\end{definition}

\textbf{RC Algorithm (\citet{bareinboim2022recovering})}\label{rc}

For \(W,Z\subseteq M\), consider the problem of recovering \(P(W| Z)\) from \(P(T)\) and \(P(M| S=1)\), and define procedure \(\mathrm{RC}(W,Z)\) as follows:

\begin{enumerate}
  \item If \(W\cup Z\subseteq T\), then \(P(W| Z)\) is s-recoverable.
  \item If 
    \[
      S \perp\!\!\!\perp W \;\big|\; Z,
    \]
    then \(P(W| Z)\) is s-recoverable as
    \[
      P(W| Z)
      = P(W| Z,\,S=1).
    \]
  \item For minimal \(C\subseteq M\) such that 
    \(\displaystyle S \perp\!\!\!\perp W \;\big|\; Z\cup C\),
    \[
      P(W| Z)
      = \sum_{c}P(W| Z,\,c,\,S=1)\;P(c| Z).
    \]
    If \(C\cup Z\subseteq T\), then \(P(W| Z)\) is s-recoverable. Otherwise, call \(\mathrm{RC}(C,Z)\).
  \item For some \(W'\subset W\),
    \[
      P(W| Z)
      = P\bigl(W'| W\setminus W',\,Z\bigr)\;P\bigl(W\setminus W'| Z\bigr).
    \]
    Call \(\mathrm{RC}\bigl(W',\{W\setminus W'\}\cup Z\bigr)\) 
    and \(\mathrm{RC}\bigl(W\setminus W',\,Z\bigr)\).
  \item Exit with \textbf{FAIL} (to s-recover \(P(W| Z)\)) if for a singleton \(W\), none of the above operations are applicable.
\end{enumerate}



\subsection{Lemmas and Proofs}

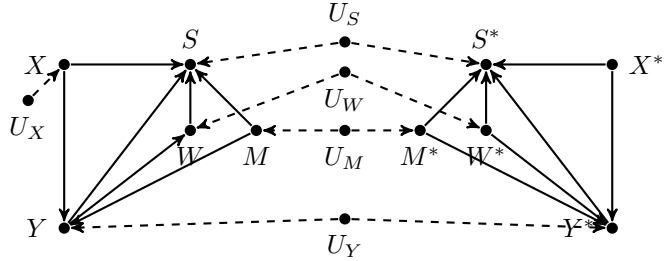
\begin{figure}[h]
\centering
\begin{tikzpicture}[->,>=stealth',node distance=2cm,
  thick,main node/.style={circle,fill,inner sep=1.5pt}]
  \node[main node] (0) [label=left:\(X\)]{};
  \node[main node] (1) [below = 2 cm of 0,label=left:\(Y\)] {};
  \node[main node] (2) [right = 1.5 cm of 0,label=above:\(S\)] {};
  \node[main node] (3) [below = 0.7 cm of 2,label=below:\(W\)] {};
  \node[main node] (4) [right = 0.7 cm of 3,label=below:\(M\)] {};
  \node[main node] (5) [right = 2 cm of 4,label=below:\(M^*\)] {};
  \node[main node] (6) [right = 0.7 cm of 5,label=below:\(W^*\)] {};
  \node[main node] (7) [above = 0.7 cm of 6,label=above:\(S^*\)] {};
  \node[main node] (8) [right = 1.5 cm of 7,label=right:\(X^*\)] {};
  \node[main node] (9) [below = 2 cm of 8,label=left:\(Y^*\)] {};
  \node[main node] (10) [right = 1 cm of 4,label=below:\(U_{M}\)] {};
  \node[main node] (11) [above = 1 cm of 10,label=above:\(U_{S}\)] {};
  \node[main node] (12) [above = 0.6 cm of 10,label=below:\(U_{W}\)] {};
  \node[main node] (13) [below = 1 cm of 10,label=below:\(U_{Y}\)] {};
  \node[main node] (14) [below left = 0.5 cm of 0,label=below:\(U_{X}\)] {};
  \path[every node/.style={font=\sffamily\small}]
    (0) edge node {} (1)
    (0) edge node {} (2)
    (1) edge node {} (2)
    (1) edge node {} (3)
    (3) edge node {} (2)
    (4) edge node {} (2)
    (4) edge node {} (1)    
    (8) edge node {} (9)
    (8) edge node {} (7)
    (9) edge node {} (7) 
    (6) edge node {} (7) 
    (6) edge node {} (9)
    (5) edge node {} (7) 
    (5) edge node {} (9);
  \draw [dashed] (11) -- (7);
  \draw [dashed] (11) -- (2);
  \draw [dashed] (12) -- (6);
  \draw [dashed] (12) -- (3);
  \draw [dashed] (10) -- (4);
  \draw [dashed] (10) -- (5);
  \draw [dashed] (13) -- (1);
  \draw [dashed] (13) -- (9);
  \draw [dashed] (14) -- (0);
\end{tikzpicture}

\caption{There are direct path, indirect path, and spurious path between \(Y\) and \(S\).}
\label{causalg_extra}
\end{figure}
\textbf{Lemma 2.}
If experimental distribution \(P(Y^*_{X^*})\) in \(G_s\) is not s-recoverable, then \(P(Y^*_{X^*})\) is not s-recoverable in the graph \(G_s^{'}\) resulting from adding a single edge to \(G_s\).

\emph{Proof.} Suppose that in the original selection‐augmented graph $G_s$, the experimental distribution $P(Y^*_{X^*})$ is not s‑recoverable.  Then there must exist two structural causal models $M_1$ and $M_2$ such that, under the biased experimental distribution given $S=1$,

\begin{itemize}
    \item $P_{G_s}^{M_1}(\mathcal{M}| S=1) = P_{G_s}^{M_2}(\mathcal{M}| S=1)$
    \item $P^{M_1}(Y^*_{X^*}) \neq P^{M_2}(Y^*_{X^*})$
\end{itemize}

Now construct a new augmented graph $G'_s$ by adding a single directed edge to the original graph $G_s$. We will show that the new graph $G'_s$ remains non-experimentally s-recoverable. Specifically, we set the parameters associated with the newly added edge to zero, effectively neutralizing this edge. Consequently, we can retain exactly the same structural models $M_1$ and $M_2$ from $G_s$ in the new graph $G'_s$, maintaining that

$$
P_{G'_s}^{M_1}(\mathcal{M}| S=1) = P_{G'_s}^{M_2}(\mathcal{M}| S=1)\quad \text{and}\quad P_{G'_s}^{M_1}(Y^*_{X^*}) \neq P_{G'_s}^{M_2}(Y^*_{X^*}).
$$

This establishes that adding a single edge to a graph that is not experimentally s-recoverable cannot render the new graph experimentally s-recoverable.

\begin{lemma}{If \(P(Y^*_{X^*})\) is naturally experimental s-recoverable, no direct and indirect path exists between the \(S\) and \(Y\) nodes in the corresponding \(G_s\).}
\end{lemma}
\emph{Proof:} Since direct and indirect paths are practically equivalent in this problem. Indirect paths can be considered as a subdivision of direct paths. I will only numerically prove the case related to direct paths here, and will provide the proof and analysis based on d-separation later.

Consider the subgraph \(G_s\) of Figure~\ref{causalg_extra} consisting only of \((S,X,Y)\). Now construct the graph \(G'_s\) to set the parameter of the path pointing from \(S\) to \(Y\) to 0. Now consider the distribution \(P_1\) that is compatible with \(G_s\), and the distribution \(P_2\) that is compatible with \(G'_s\). and make \(P_1(Y^*_{X^*} | S=1) = P_2(Y^*_{X^*})\). 

Assume $Y_x\in\{0,1\}$ with $P(Y^*_x=1)=P(Y^*_x=0)=\tfrac12$ in the unbiased population.  Define
\[
\alpha \;=\; P(S=1 | Y^*_x=1), 
\quad
\beta \;=\; P(S=1 | Y^*_x=0),
\]
and suppose $0<\alpha<\beta<1$. 

By Bayes’ rule,
\[
P_2(Y^*_x) = P_1(Y^*_x | S=1)
= \frac{P(S=1 | Y^*_x)\,P(Y^*_x)}{P(S=1)}
= \frac{P(S=1 | Y^*_x)\,P(Y^*_x)}{\sum_{y}P(S=1| Y^*_x=y)\,P(Y^*_x=y)}
\]
\[
P_1(Y^*_x=1 | S=1)
= \frac{\alpha \cdot \tfrac12}{\alpha \cdot \tfrac12 + \beta \cdot \tfrac12}
= \frac{\alpha}{\alpha + \beta}
\;\neq\;\tfrac12
\]

Hence $P_1(Y^*_x| S=1)\neq P_2(Y^*_x)$, and the model with $Y_x\!\to\!S$ is \emph{not} naturally experimental s‑recoverable.

\emph{Analysis:} Consider Figure~\ref{causalg_extra}, where both direct and indirect paths exist between nodes \( S \) and \( Y \). By constructing a twin network through shared exogenous variables, it becomes clear that node \( S \) can directly affect the counterfactual node \( Y^*_{X^*} \) in the counterfactual scenario. Consequently, the presence of either direct or indirect paths connecting nodes \( S \) and \( Y \) disrupts the natural s-recoverability of the corresponding distribution \( P(Y^*_{X^*}) \). For instance, if the selection criterion for data collection explicitly depends upon the experimental outcomes, unbiased estimation of causal effects becomes inherently impossible.

Notably, the scenario involving spurious paths between nodes \( S \) and \( Y \) is more nuanced. Consider Figure~\ref{causalg_c}: although a spurious pathway connects \( S \) and \( Y \), the distribution \( P(Y^*_{X^*}) \) compatible with Figure~\ref{causalg_c} still satisfies natural experimental s-recoverability due to the intervention on the counterfactual variable \( X \). Conversely, in Figure~\ref{causalg_extra}, the confounder \( M \) propagates selection bias toward \( Y^*_{X^*} \) via node \( Y \) and its corresponding exogenous variable \( U_Y \). Therefore, no straightforward criterion exists to determine whether a spurious path between \( S \) and \( Y \) necessarily violates natural s-recoverability.


\textbf{Theorem 2.}
Let \(G_s\) be a causal graph augmented with a selection node \(S\), and let \(V\) denote the set of all variables. Suppose there exists a subset \(Z \subseteq V\) that is measured in both the biased experiment and at the population level, and that \((Y^*_{X^*} \perp\!\!\!\perp S | Z.\)) Then, the experimental distribution is s-recoverable: \(P(Y^*_{X^*}) \;=\; \sum_{z} P(Y^*_{X^*} | Z, S=1)\, P(Z).\)

\emph{Proof :}  We can condition on set \(Z\):
\[
\begin{aligned}
P(Y^*_{X^*} )
& = \sum_{z} P(Y^*_{X^*} | Z)\, P(Z) \\
& = \sum_{z} P(Y^*_{X^*} | Z, S=1)\, P(Z) 
 \end{aligned}
\]
Where the last equation follows that \(Y^* \perp\!\!\!\perp \ S | Z\).

\subsection{Experimental and Computational Details}

\subsubsection{computational detail of discrete experiment}
Here is the computational detail of experiment \ref{exp_dis}.

\[
\begin{aligned}
P(Y^*_{x^*})
&= P\bigl(Y | do(X = x)\bigr) \\
&= \sum_{w\in\{0,1\}} P\bigl(Y | do(X = x), W = w\bigr)\,P(W = w) \\
&= \sum_{w\in\{0,1\}} P\bigl(Y | X = x, W = w\bigr)\,P(W = w) \\
&\implies
\begin{cases}
P(Y^*_{x^*=1} = 1)
= \tfrac12\Bigl[\tfrac{0.95 + 0.80}{2} \;+\;\tfrac{0.90 + 0.60}{2}\Bigr]
= 0.8125\\[6pt]
P(Y^*_{x^*=1} = 0) = 1- P(Y^*_{x^*=1} = 1) = 0.1875\\[6pt]
P(Y^*_{x^*=0} = 1) \tfrac12\Bigl[\tfrac{0.90 + 0.50}{2} \;+\;\tfrac{0.70 + 0.30}{2}\Bigr]= 0.60 \\  
P(Y^*_{x^*=0} = 0) = 1- P(Y^*_{x^*=0} = 1) = 0.40\\
\end{cases}
\end{aligned}
\]

By Theorem~\ref{thm2} . We only need external distribution for \(Z\) to restore the biased experimental distribution to an unbiased one. From the open source dataset, we know that \(Z\sim\mathrm{Bernoulli}(0.5)\).

\[
\begin{aligned}
P_{rec}(Y^*_{x^*})
&= \sum_{z} P\bigl(Y^*_{x^*} | Z=z,\,S=1\bigr)\,P(Z=z)\\
&\implies
\begin{cases}
P(Y^*_{x^*=1} = 1)
= \tfrac12\Bigl[\tfrac{158+146}{158+146+8+10} \;+\;\tfrac{266+218}{266+218+73+146}\Bigr]
\approx 0.816\\[6pt]
P(Y^*_{x^*=0} = 1)
= \tfrac12\Bigl[\tfrac{141+100}{141+100+12+245} \;+\;\tfrac{180+110}{180+110+174+245}\Bigr]
\approx 0.613\\
P(Y^*_{x^*=1} = 0) 
=1-P(Y^*_{x^*=1} = 1)
\approx 1-0.816=0.184\\
P(Y^*_{x^*=0} = 0) 
=1-P(Y^*_{x^*=0} = 1)
\approx 1-0.613 = 0.387
\end{cases}
\end{aligned}
\]

By experimental data, it is easy to obtain:
\[
\begin{aligned}
P\bigl(Y^*_{x^*=0}=1 | S=1\bigr)
&= \frac{531}{531 + 473}
\approx 0.529\\
P\bigl(Y^*_{x^*=0}=0 | S=1\bigr)
&= 1-0.529
=0.471\\
P\bigl(Y^*_{x^*=1}=1 | S=1\bigr)
&= \frac{788}{788 + 237}
\approx 0.768\\
P\bigl(Y^*_{x^*=1}=0 | S=1\bigr)
&= 1-0.768
=0.232
\end{aligned}
\]

The formula for calculating relative error is as follows:

\[
\begin{aligned}
\mathrm{RE}_{\mathrm{bias}}(x)
&= \frac{ P_{\mathrm{bias}}(Y^*_{x^*}=1 | S=1) - P_{\mathrm{true}}(Y^*_{x^*}=1)}
       {P_{\mathrm{true}}(Y^*_{x^*}=1)},\\
\mathrm{RE}_{\mathrm{rec}}(x)
&= \frac{\hat P_{\mathrm{rec}}(Y^*_{x^*}=1) - P_{\mathrm{true}}(Y^*_{x^*}=1)}
       {P_{\mathrm{true}}(Y^*_{x^*}=1)}.
\end{aligned}
\]

\subsubsection{Computational detail of continuous experiment}

We provide the full derivation of the experimental distribution \( P(Y_x) \) under a linear structural causal model (SCM) with Gaussian noise.

\paragraph{Structural Equations.}  
Let the SCM be defined as:
\[
Y = \alpha X + \beta W + \gamma Z + U_Y
\]
where \( \alpha, \beta, \gamma \in \mathbb{R} \) are fixed coefficients, and \( U_Y \sim \mathcal{N}(0, \sigma_Y^2) \) is an independent exogenous variable term.  
We assume the covariates:
\[
W \sim \mathcal{N}(0, \sigma_W^2), \quad Z \sim \mathcal{N}(0, \sigma_Z^2), \quad W \perp Z \perp U_Y
\]

\paragraph{Intervention.}  
To compute the experimental distribution \( P(Y_x) \), we apply the do-operator \( do(X = x) \), which modifies the SCM by setting \( X = x \) and removing any edges into \( X \).  
The structural equation becomes:
\[
Y^*_{X^*=x} = \alpha x + \beta W + \gamma Z + U_Y
\]

\paragraph{Distribution of $Y^*_{X^*=x}$}  
We now compute the distribution of \( Y^*_{X^*=x} \) by leveraging the independence and Gaussianity of \( W, Z, U_Y \). Since \( Y^*_{X^*=x} \) is a linear combination of independent Gaussian variables, it is also Gaussian:
\[
Y^*_{X^*=x} \sim \mathcal{N}(\mu, \sigma^2)
\]
We compute the mean:
\begin{align*}
\mathbb{E}[Y^*_{X^*=x}] &= \mathbb{E}[\alpha x + \beta W + \gamma Z + U_Y] \\
                &= \alpha x + \beta \mathbb{E}[W] + \gamma \mathbb{E}[Z] + \mathbb{E}[U_Y] \\
                &= \alpha x
\end{align*}
since \( \mathbb{E}[W] = \mathbb{E}[Z] = \mathbb{E}[U_Y] = 0 \).  

The variance is:
\begin{align*}
\mathrm{Var}(Y_x) &= \mathrm{Var}(\beta W + \gamma Z + U_Y) \\
                  &= \beta^2 \mathrm{Var}(W) + \gamma^2 \mathrm{Var}(Z) + \mathrm{Var}(U_Y) \\
                  &= \beta^2 \sigma_W^2 + \gamma^2 \sigma_Z^2 + \sigma_Y^2
\end{align*}

\paragraph{Conclusion.}  
Thus, the experimental distribution \( Y^*_{X^*=x} \) is:
\[
Y^*_{X^*=x} = \mathcal{N}\left(\alpha x,\ \beta^2 \sigma_W^2 + \gamma^2 \sigma_Z^2 + \sigma_Y^2\right)
\]
In our continuous experiment where \( \sigma_W^2 = \sigma_Z^2 = 1 \), this simplifies to:
\[
Y^*_{X^*=x} = \mathcal{N}\left(\alpha x,\ \beta^2 + \gamma^2 + \sigma_Y^2\right)
\]

As long as the above SCM model is met, the theoretical experimental distribution can be calculated by simply bringing in different parameters.

\subsection{Algorithm failure analysis}

\label{failureexample}
\begin{figure}[h]
\centering
\begin{tikzpicture}[->,>=stealth',node distance=2cm,
  thick,main node/.style={circle,fill,inner sep=1.5pt}]

  \node[main node] (0) [label=below left:{\(X\)}]{};
  \node[main node] (1) [below =2.5cm of 0,label=left:\(Y\)] {};

  \node[main node] (3) [below left =1cm of 1,label=left:\(W_2\)] {};
  \node[main node] (4) [above right =1.5cm of 0,label=left:\(S\)] {};
  \node[main node] (5) [below =1.5cm of 4,label=left:\(W_3\)] {};
  \node[main node] (6) [below =1cm of 5,label=left:\(W_4\)] {};  
  \node[main node] (7) [right =4cm of 0, label=above:{\(X^*\)}]{};
  \node[main node] (8) [below =2.5cm of 7,label=right:\(Y^*\)] {};
  \node[main node] (9) [above left =1.5cm of 7,label=right:\(S^*\)] {};
  \node[main node] (10) [below =1.5cm of 9,label=right:\(W_3^*\)] {};
  \node[main node] (11) [below =1cm of 10,label=right:\(W_4^*\)] {};

  \node[main node] (13) [below right =1cm of 8,label=right:\(W_2^*\)] {};
  \node[main node] (14) [right =1.95cm of 1,label=below:\(U_Y\)] {};  
  \node[main node] (15) [above =0.7cm of 14,label=below:\(U_{W_4}\)] {}; 
  \node[main node] (16) [above =1.7cm of 14,label=below:\(U_{W_3}\)] {};
  \node[main node] (17) [below =0.65cm of 14,label=below:\(U_{W_2}\)] {};
  \node[main node] (18) [above =3cm of 14,label=below:\(U_{S}\)] {};
  \node[main node] (19) [left =1.9cm of 4,label=left:\(W_1\)] {};
  \node[main node] (20) [right =1.8cm of 9,label=right:\(W_1^*\)] {};
  \node[main node] (21) [above =4.5cm of 14,label=below:\(U_{W_1}\)] {};
  \node[main node] (22) [above =0.55cm of 0,label=above:\(U_{X}\)] {};  

  \path[every node/.style={font=\sffamily\small}]
    (0) edge node {} (1)
    (7) edge node {} (8)
    (0) edge node {} (19)    
    (19) edge node {} (3)
    (20) edge node {} (13)
    (3) edge node {} (1)
    (13) edge node {} (8)
    (3) edge node {} (1)
    (13) edge node {} (8)
    (0) edge node {} (4)
    (7) edge node {} (9)    
    (5) edge node {} (4)
    (10) edge node {} (9)
    (6) edge node {} (5)
    (11) edge node {} (10)
    (6) edge node {} (1)
    (3) edge node {} (4)
    (7) edge node {} (20)
    (13) edge node {} (9)
    (11) edge node {} (8);
  \draw [dashed] (18) -- (9);
  \draw [dashed] (18) -- (4);
  \draw [dashed] (14) -- (8);
  \draw [dashed] (14) -- (1);
  \draw [dashed] (15) -- (11);
  \draw [dashed] (15) -- (6); 
  \draw [dashed] (16) -- (10);
  \draw [dashed] (16) -- (5); 
  \draw [dashed] (17) -- (3); 
  \draw [dashed] (17) -- (13); 
  \draw [dashed] (21) -- (19); 
  \draw [dashed] (21) -- (20); 
  \draw [dashed] (22) -- (0);   
\end{tikzpicture}
\caption{}
\label{causalg_extra_2}
\end{figure}

Using Algorithm~\ref{alg:twin}, we can conduct a more systematic analysis of whether the distribution \(P(Y^*_{x^*})\) is s-recoverable. For instance, when unbiased distributions for certain nodes are not readily available, the current algorithm may fail to yield a simple set of variables that guarantees the experimemtal s-recoverability of \(P(Y^*_{x^*})\). In such cases, one may employ a recursive procedure to recover the unbiased distributions for these nodes, or even resort to a chain rule factorization of conditional probabilities in order to identify a more complex yet effective solution. The core principle of the algorithm remains rooted in the notion of d-separation.
\begin{lemma}{Algorithm~\ref{alg:twin} does not guarantee a valid output on all graphs.}

There exist graphs for which no set of variables can d-separate \(S\) and \(Y^*_{x^*}\). Consequently, the algorithm is not universally applicable, and it cannot guarantee a valid solution for every graph. Consider, for example, Figure~\ref{causalg_extra_2}. In the corresponding twin network, there exists the following path:
\[
S \leftarrow W_2 \leftarrow U_{W_2} \rightarrow W_2^* \rightarrow Y^*,
\]
which necessitates conditioning on \(W_2\) to block the influence of \(S\) on \(Y'\). Unfortunately, conditioning on \(W_2\) simultaneously opens up a previously blocked path:
\[
S \leftarrow X \rightarrow W_1 \rightarrow W_2 \leftarrow U_{W_2} \rightarrow W_2^* \rightarrow Y^*,
\]
since \(W_2\) functions as a collider on this path. Alternatively, if one attempts to condition on \(W_1\) to block this route, a new path is activated:
\[
S \leftarrow X \rightarrow W_1 \leftarrow U_{W_1} \rightarrow W_1^* \rightarrow W_2^* \rightarrow Y^*,
\]
Therefore, this algorithm fails on this causal graph, and The algorithm does not guarantee a valid output on all graphs.
\end{lemma}

\subsection{Supplementary Experiments}
\subsubsection{Supplementary figures in continuous experiment}

\begin{figure}[H]        
  \centering
  \includegraphics[width=1.0\linewidth]{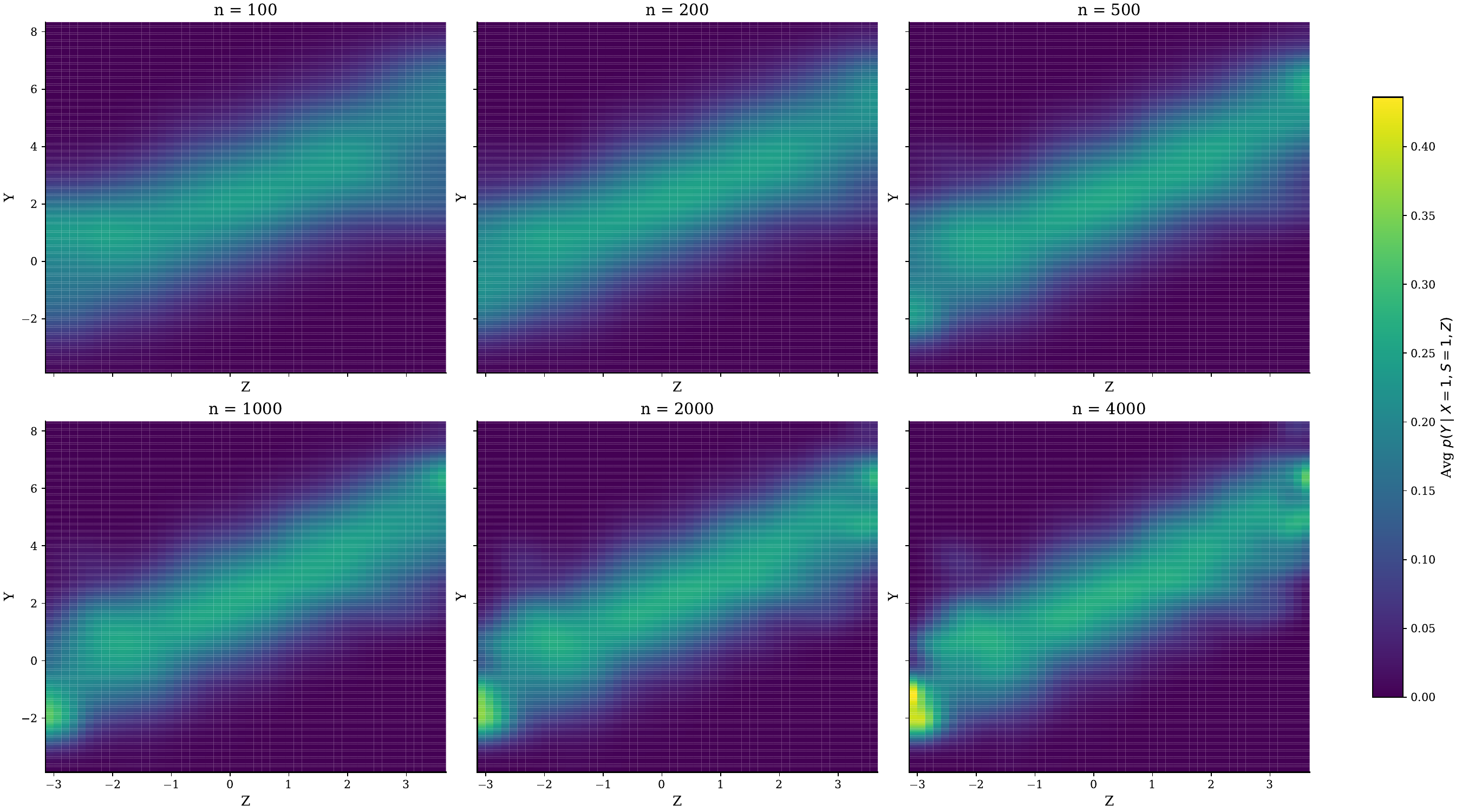}
  \caption{Kernel density estimates of \(P\bigl(Y^*_{X^*=1}=1| Z, S=1\bigr)\) obtained via 50 independent random seeds at each sample size.}
  \label{fig:density}
\end{figure}

\begin{figure}[H]        
  \centering
  \makebox[\textwidth][c]{%
  \includegraphics[width=1.0\linewidth]{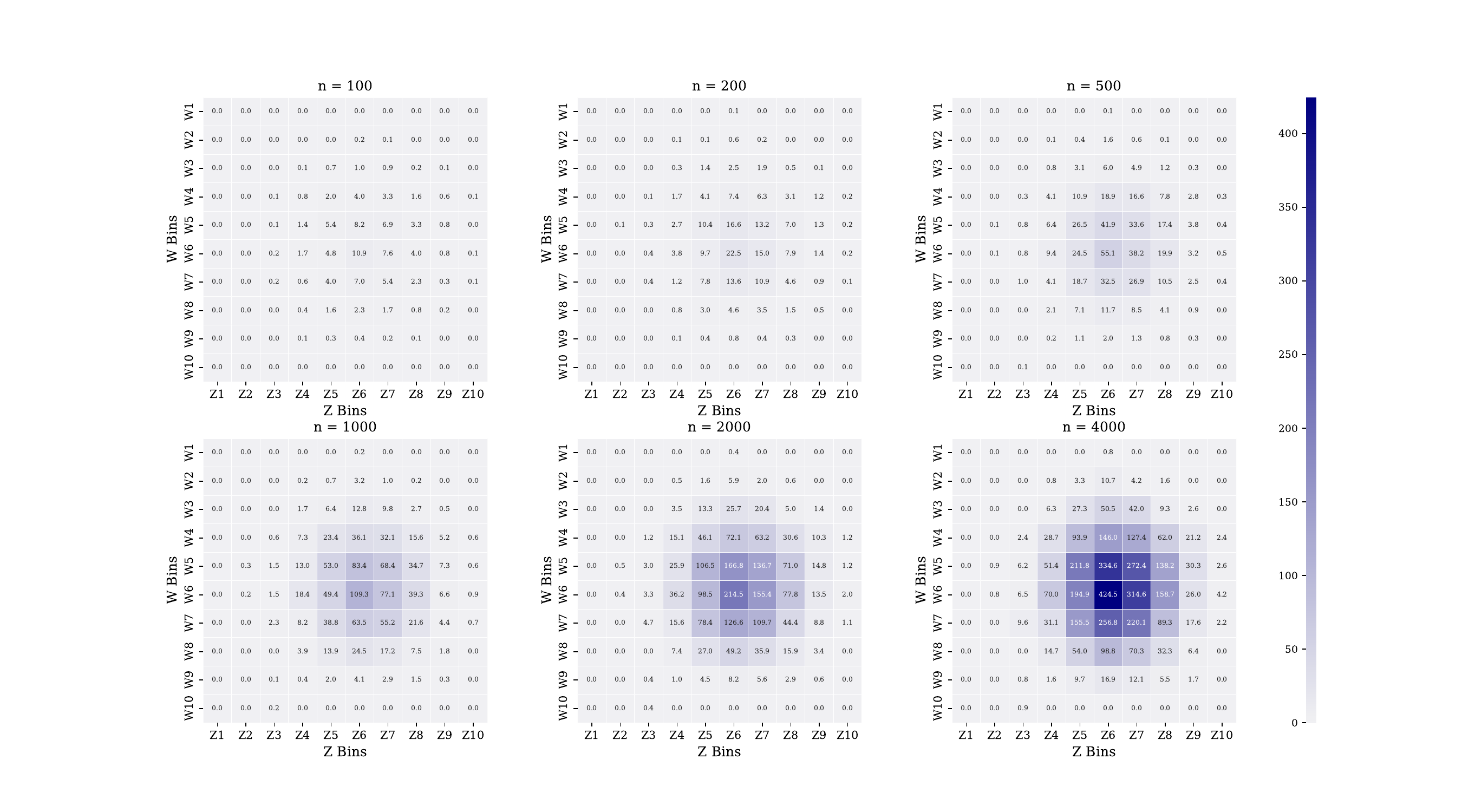}
  }
  \caption{Counts information in each \((w,c)\) cells in biased continuous example.}
  \label{fig:counts}
\end{figure}

Figure~\ref{fig:counts} shows the counts in each \((W,Z)\) bin under selection bias for sample sizes \(n\in\{100,200,500,1000,2000,4000\}\).  For small samples (\(n=100,200\)), most bins record zero observations, with only a few central bins containing minimal counts.  At medium sample sizes (\(n=500,1000\)), central bins rise to the tens, while peripheral bins remain sparse.  At large sample sizes (\(n=2000,4000\)), central bins accumulate counts in the hundreds, and even moderate‑probability bins reach tens of observations, providing sufficient support for KDE.  The evolution of these counts corresponds directly to the KDE estimates’ convergence from high noise to smooth accuracy, highlighting that, under selection bias, adequate coverage of the covariate space is critical for recovering conditional distributions.
\subsubsection{Advanced continuous example}\label{ace}

Note: All experiments conducted in this paper can be reproduced on PC and Linux systems with no computational resource requirements.

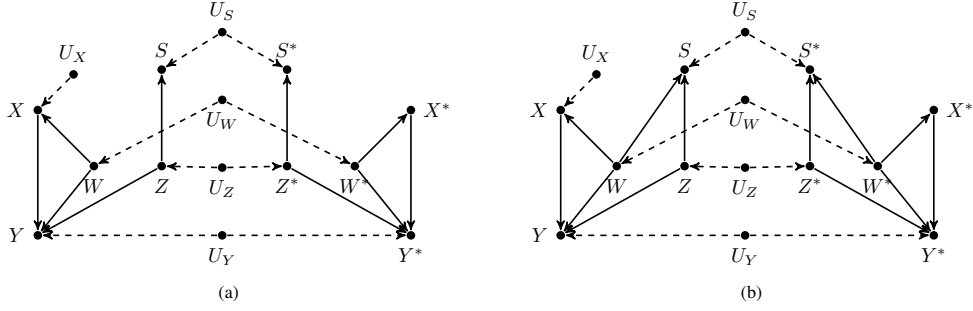
\begin{figure}[t]
\centering
\resizebox{1.0\textwidth}{!}{%
    \begin{subfigure}[b]{0.65\textwidth}
    \centering
        \begin{tikzpicture}[->,>=stealth',node distance=2cm,
      thick,main node/.style={circle,fill,inner sep=1.5pt}]
      
      \node[main node] (0) [label=left:\(X\)]{};
      \node[main node] (1) [below = 2 cm of 0,label=left:\(Y\)] {};
      \node[main node] (2) [below right = 1.2 cm of 0,label=below:\(W\)] {}; 
      \node[main node] (3) [right = 1.0 cm of 2,label=below:\(Z\)] {}; 
      \node[main node] (4) [above = 1.5 cm of 3,label=above:\(S\)] {};
      \node[main node] (5) [right = 2 cm of 4,label=above:\(S^*\)] {};
      \node[main node] (6) [below = 1.5 cm of 5, label=below:\(Z^*\)] {};
      \node[main node] (7) [right = 1.0 cm of 6, label=below:\(W^*\)] {};
      \node[main node] (8) [above right = 1.2 cm of 7, label=right:\(X^*\)] {};
      \node[main node] (9) [below = 2.0 cm of 8, label=below:\(Y^*\)] {};
      \node[main node] (10) [left = 3.1 cm of 9,label=below:\(U_{Y}\)] {}; 
      \node[main node] (11) [above = 1 cm of 10,label=below:\(U_{Z}\)] {};
      \node[main node] (12) [above = 1 cm of 11,label=below:\(U_{W}\)] {};
      \node[main node] (13) [above = 1 cm of 12,label=above:\(U_{S}\)] {};
      \node[main node] (14) [above right = 0.7 cm of 0,label=above:\(U_{X}\)] {}; 
      \path[every node/.style={font=\sffamily\small}]

        (0) edge node {} (1)
        (8) edge node {} (9)    
        (2) edge node {} (0)
        (2) edge node {} (1)    
        (7) edge node {} (8)
        (7) edge node {} (9)
        (3) edge node {} (1)
        (3) edge node {} (4) 
        (6) edge node {} (5)    
        (6) edge node {} (9);
      \draw [dashed] (10) -- (1);
      \draw [dashed] (10) -- (9);
      \draw [dashed] (11) -- (3);
      \draw [dashed] (11) -- (6);
      \draw [dashed] (12) -- (2);
      \draw [dashed] (12) -- (7); 
      \draw [dashed] (13) -- (4);
      \draw [dashed] (13) -- (5); 
      \draw [dashed] (14) -- (0); 
    \end{tikzpicture}
    \caption{}
    \label{causalg_ap_a}
    \end{subfigure}\hfill
    \begin{subfigure}[b]{0.65\textwidth}
    \centering
        \begin{tikzpicture}[->,>=stealth',node distance=2cm,
      thick,main node/.style={circle,fill,inner sep=1.5pt}]
      
      \node[main node] (0) [label=left:\(X\)]{};
      \node[main node] (1) [below = 2 cm of 0,label=left:\(Y\)] {};
      \node[main node] (2) [below right = 1.2 cm of 0,label=below:\(W\)] {}; 
      \node[main node] (3) [right = 1.0 cm of 2,label=below:\(Z\)] {}; 
      \node[main node] (4) [above = 1.5 cm of 3,label=above:\(S\)] {};
      \node[main node] (5) [right = 2 cm of 4,label=above:\(S^*\)] {};
      \node[main node] (6) [below = 1.5 cm of 5, label=below:\(Z^*\)] {};
      \node[main node] (7) [right = 1.0 cm of 6, label=below:\(W^*\)] {};
      \node[main node] (8) [above right = 1.2 cm of 7, label=right:\(X^*\)] {};
      \node[main node] (9) [below = 2.0 cm of 8, label=below:\(Y^*\)] {};
      \node[main node] (10) [left = 3.1 cm of 9,label=below:\(U_{Y}\)] {}; 
      \node[main node] (11) [above = 1 cm of 10,label=below:\(U_{Z}\)] {};
      \node[main node] (12) [above = 1 cm of 11,label=below:\(U_{W}\)] {};
      \node[main node] (13) [above = 1 cm of 12,label=above:\(U_{S}\)] {};
      \node[main node] (14) [above right = 0.7 cm of 0,label=above:\(U_{X}\)] {}; 
      \path[every node/.style={font=\sffamily\small}]

        (0) edge node {} (1)
        (8) edge node {} (9)    
        (2) edge node {} (0)
        (2) edge node {} (1)    
        (7) edge node {} (8)
        (7) edge node {} (9)
        (3) edge node {} (1)
        (3) edge node {} (4) 
        (6) edge node {} (5)
        (2) edge node {} (4) 
        (7) edge node {} (5)
        (6) edge node {} (9);
      \draw [dashed] (10) -- (1);
      \draw [dashed] (10) -- (9);
      \draw [dashed] (11) -- (3);
      \draw [dashed] (11) -- (6);
      \draw [dashed] (12) -- (2);
      \draw [dashed] (12) -- (7); 
      \draw [dashed] (13) -- (4);
      \draw [dashed] (13) -- (5); 
      \draw [dashed] (14) -- (0); 
    \end{tikzpicture}
    \caption{}
    \label{causalg_ap_b}
    \end{subfigure}
}
\caption{Both Figures (a) and (b) can satisfy the experimental s-recoverability by partial external data. 
}
\label{fig_ap}
\end{figure}

Follow the continuous exmaple in section~\ref{continuous_example}, we simulate a clinical trial designed to evaluate a novel therapy for a specific pulmonary condition. Participants are recruited based on their baseline inflammatory biomarker levels, denoted by \(Z\), and latent health level, denoted by \(W\). Once enrolled, treatment assignment \(X\) (novel drug:X=1 vs. standard care:X=0) is randomized via a Bernoulli draw.
The researchers have updated their selection policy to also take patients’ latent health status \(Z\) into account when recruiting patients; consequently, we obtain a new generative mechanism for \(S\):
\[
S \;=\; \mathbf{1}\bigl\{\;\gamma_W\,W \;+\;\gamma_Z\,Z \;+\; U_S > c\}\,, 
\quad U_S \sim \mathcal{N}(0,\sigma_S^2).
\]
All other parts of the SCM remain unchanged (Figure~\ref{causalg_ap_b} shows the corresponding causal diagrams). In this experiment, we furthermore make no assumptions about the SCM’s functional form or the distributions of its exogenous variables, thereby stress‐testing our estimator’s ability to recover \(P(Y^*_{x^*})\) purely from data in the absence of any prior structural knowledge.

We adopted the same experimental logic as for the experiments in section~\ref{continuous_example}. We assume that investigators can collect biased experimental cohorts of sizes \(n\in\{100,200,500,1000,2000,4000\}\).  For each \(n\), we draw 50 independent samples (using distinct random seeds) from the full synthetic dataset, computing and recording the average recovered experimental distribution \(\hat P_{\mathrm{rec}}(Y^*_{x^*})\), its average error relative to the ground truth, and the average biased experimental distribution \(\hat P_{\mathrm{bias}}(Y^*_{x^*})\).
For all simulations, we fix \(N=20000\), \(\alpha=2.0\), \(\beta=1.0\), \(\gamma_{WY}=1.0\), \(\sigma_Y=1.0\), \(\gamma_Z=0.5\), \(\gamma_W=0.5\), \(\sigma_S=1.0\), \(c=0.2\), and \(p_X=0.5\).

\begin{figure}[h]        
  \centering
  \includegraphics[width=1\linewidth]{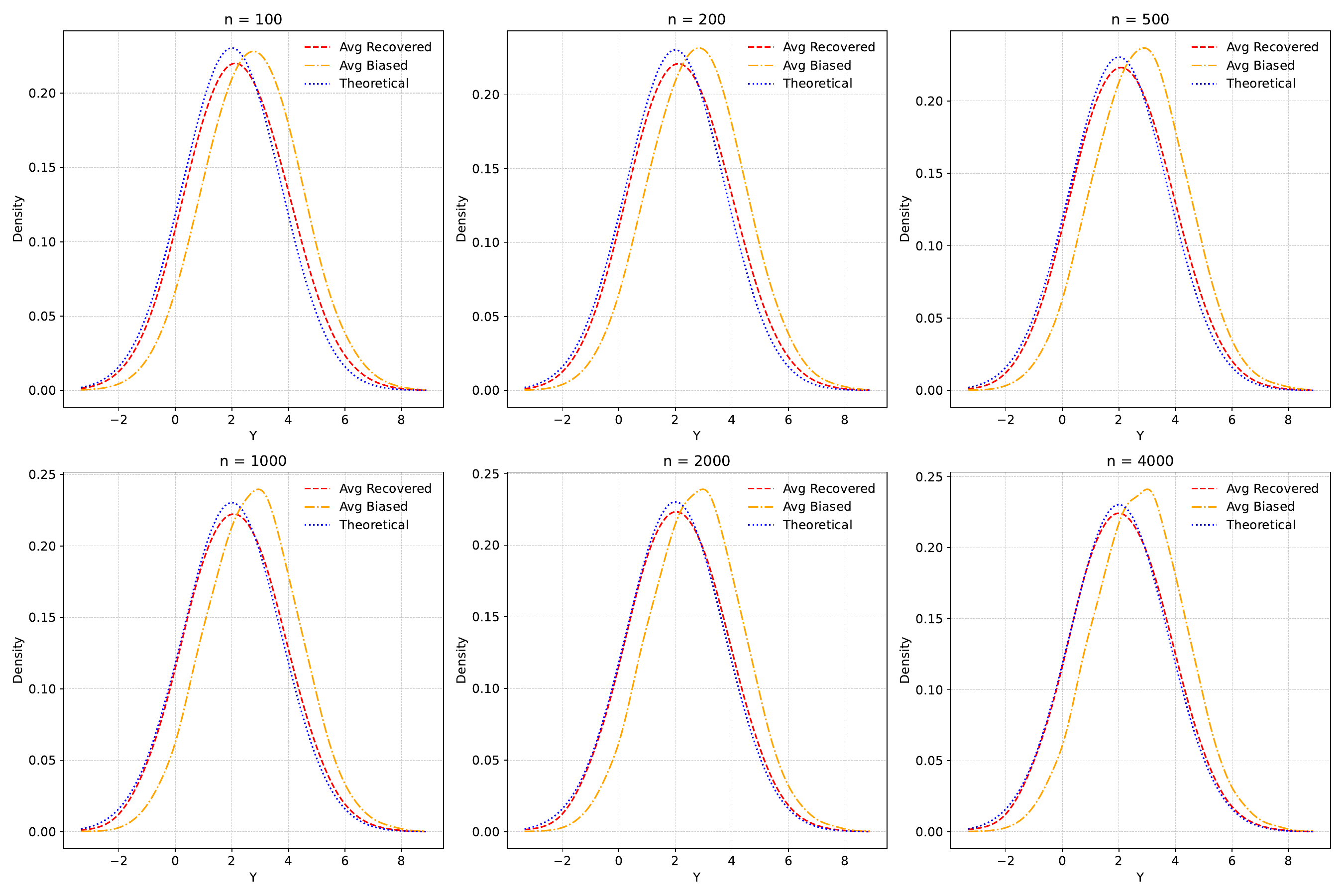}
  \caption{Density comparison of average recovered \(\overline{P}(Y^*_{x^*})\) of advanced version, average conditional \(\overline{P}(Y^*_{x^*}| S=1)\), and theoretical \(P(Y^*_{x^*})\) for sample sizes \(n\in\{100,200,500,1000,2000,4000\}\).}
  \label{fig:myplot}
\end{figure}
\begin{figure}[H]        
  \centering
  \includegraphics[width=1.0\linewidth]{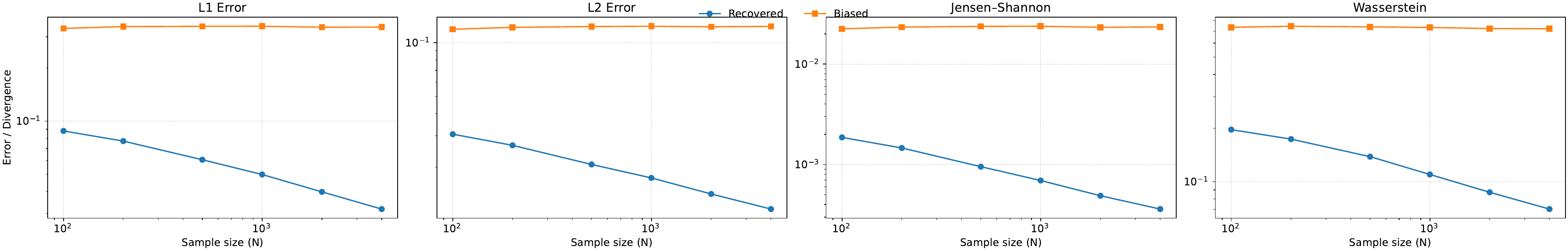}
  \caption{Comparison of error metrics of advanced version between the recovered experimental distribution and the biased follow‑up distribution across sample sizes \(N\).Figures (a)–(d) display, respectively, (a) L1 error, (b) L2 error, (c) Jensen–Shannon divergence, and (d) Wasserstein distance, averaged over 50 random seeds. }
  \label{fig:ad_error}
\end{figure}
\begin{table}[H]
\centering
\caption{Error metrics comparing recovered and biased distributions.}
\label{tab:error_metrics}
\resizebox{0.95\textwidth}{!}{%
    \begin{tabular}{rrrrrrrrr}
    \toprule
    N & L1\_rec & L1\_ias & L2\_rec & L2\_bias & JS\_rec & JS\_bias & Wass\_rec & Wass\_bias \\
    |rule
    100 & 0.0880 & 0.3346 & 0.0306 & 0.1191 & 0.0019 & 0.0224 & 0.1961 & 0.7357 \\
    200 & 0.0771 & 0.3428 & 0.0265 & 0.1221 & 0.0015 & 0.0233 & 0.1735 & 0.7458 \\
    500 & 0.0605 & 0.3439 & 0.0207 & 0.1234 & 0.0010 & 0.0237 & 0.1383 & 0.7400 \\
    1000 & 0.0499 & 0.3442 & 0.0174 & 0.1238 & 0.0007 & 0.0238 & 0.1099 & 0.7345 \\
    2000 & 0.0398 & 0.3399 & 0.0141 & 0.1229 & 0.0005 & 0.0232 & 0.0873 & 0.7237 \\
    4000 & 0.0318 & 0.3404 & 0.0116 & 0.1237 & 0.0004 & 0.0234 & 0.0703 & 0.7224 \\
    \bottomrule
    \end{tabular}%
}
\end{table}

The experiments validate the efficacy of our proposed nonparametric approach for correcting experimental distributions distorted by complex selection mechanisms. Although the selection indicator $S$ depends jointly on variables $W$ and $Z$, inducing significant systematic bias, our method leverages externally available unbiased marginal distributions $P(W)$ and $P(Z)$ to reweight and integrate the conditional density. The recovered experimental distribution $P(Y^*_{x^*})$ significantly outperforms the original biased distribution across multiple error metrics, including L1, L2, Jensen–Shannon divergence, and Wasserstein distance, and rapidly converges toward the theoretical distribution as the sample size increases. This result demonstrates that our approach achieves robustness and consistency without relying on structural assumptions or parametric models, thus providing a reliable and broadly applicable method for experimental distribution correction in practical causal inference settings.

\subsection{Discussion}
It is worth emphasizing that the identification of experimental distributions is not required in our recovery procedure. This principle is rigorously adhered to in both our definitions and algorithms, as we avoid explicitly converting the experimental distribution $P(Y^*_{x^*})$ into the form $P(y | do(x))$ and subsequently attempting identification. Such conversions often introduce complex and intertwined problems of identification and estimation from observational distributions. Instead, our objective remains strictly to recover the unbiased distribution $P(Y^*_{x^*})$ directly from the available biased experimental data $P(Y^*_{x^*} | S=1)$. Although the equivalence between $P(Y^*_{x^*})$ and $P(y | do(x))$ is well-established, no analogous equivalence necessarily holds between $P(Y^*_{x^*} | S=1)$ and $P(y | do(x), S=1)$. Therefore, one cannot straightforwardly reduce the task of recovering $P(Y^*_{x^*} | S=1)$ to recovering $P(y | do(x), S=1)$.

Consider, for example, an integrated approach that couples identification and recovery:

\[
\begin{aligned}
P(Y^*_{x^*}) 
&= P(y | do(x)) \\
&= \sum_z P(y | do(x), Z)P(Z | do(x)) \\
&= \sum_z P(y | do(x), Z, S=1)P(Z | do(x)) \\
&= \sum_z P(y | do(x), Z, S=1)P(Z | do(x)) \\
&= \sum_{z,w,m} P(y | do(x), Z, w, S=1)P(w | x, Z, S=1)P(Z | M, do(x))P(M | do(x)) \\
&= \sum_{z,w,m} P(y | x, Z, w, S=1)P(w | x, Z, S=1)P(Z | M, x)P(M | x).
\end{aligned}
\]

Such an approach conflates the unbiased recovery of a distribution with its identification. It relies on the equivalence of $P(Y^*_{x^*})$ and $P(y | do(x))$, and involves identifying a node set that d-separates $Y$ and $S$ in the residual graph obtained by removing incoming edges to $X$. Subsequently, the method conditions on a complex observational set W, thereby accomplishing identification. While this approach recovers $P(Y^*_{x^*})$ through a combination of biased and unbiased observational data, it faces significant drawbacks: first, identifying suitable sets $Z,W$ and $M$ is computationally intensive, dramatically increasing complexity; second, coupling recovery and identification obscures error attribution, hindering clarity in experimental analysis; third, overly complicated conditioning sets $W$ are often difficult to obtain.

In contrast, our method directly recovers the unbiased experimental distribution $P(Y^*_{x^*})$ from the biased experimental data $P(Y^*_{x^*}| S=1)$ by leveraging readily available unbiased observational data. This decoupling of recovery from the identification process not only simplifies the overall estimation procedure but also enhances both interpretability and practical applicability.




\end{document}